\documentclass[usenatbib]{mn2e}
\usepackage{epsfig}

\title{Abundance of damped Lyman-$\alpha$ absorbers in cosmological
SPH simulations}

\author[Nagamine, Springel, \& Hernquist]
  {K.~Nagamine,$^1$\thanks{Email: knagamin@cfa.harvard.edu}  
   V.~Springel,$^2$\thanks{Email: volker@mpa-garching.mpg.de}
   and L.~Hernquist~$^1$\thanks{Email: lars@cfa.harvard.edu} \vspace{0.3cm}\\
  $^1$Harvard-Smithsonian Center for Astrophysics, 
	60 Garden Street, Cambridge, MA 02138, U.S.A. \\
  $^2$Max-Planck-Institut f\"{u}r Astrophysik, 
	Karl-Schwarzschild-Stra\ss{}e 1, 85740 Garching bei 
	M\"{u}nchen, Germany}

\pagerange{\pageref{firstpage}--\pageref{lastpage}}
\pubyear{2002}

\newcommand{\Lam}{\Lambda}

\newcommand{\Del}{\Delta}

\newcommand{\cm}{\rm cm}
\newcommand{\kms}{\,\rm km\, s^{-1}}
\newcommand{\Msun}{\rm M_{\odot}}

\newcommand{\hinv}{h^{-1}}
\newcommand{\himpc}{\hinv{\rm\,Mpc}}
\newcommand{\hikpc}{\hinv{\rm\,kpc}}
\newcommand{\himsun}{\,\hinv{\Msun}}

\newcommand{\Om}{\Omega_{\rm m}}
\newcommand{\Ol}{\Omega_{\Lam}}
\newcommand{\Ob}{\Omega_{\rm b}}
\newcommand{\OHI}{\Omega_{\rm HI}}
\newcommand{\HI}{H{\sc i}\,\,}
\newcommand{\NHI}{{N_{\rm HI}}}

\newcommand{\XH}{X_{\rm H}}
\newcommand{\sdla}{\sigma_{\rm DLA}}
\newcommand{\Mtot}{{M_{\rm tot}}}

\newcommand{\beq}{\begin{eqnarray}}
\newcommand{\eeq}{\end{eqnarray}}

\begin{document}

\maketitle

\label{firstpage}


\begin{abstract}

We use cosmological smoothed-particle hydrodynamics (SPH) simulations
of the $\Lam$ cold dark matter (CDM) model to study the abundance of
damped Lyman-$\alpha$ absorbers (DLAs) in the redshift range
$z=0-4.5$.  We compute the cumulative DLA abundance by using the
relation between DLA cross-section and the total halo mass inferred
from the simulations.  Our approach includes standard radiative
cooling and heating with a uniform UV background, star formation,
supernova feedback, as well as a phenomenological model for feedback
by galactic winds. The latter allows us to examine, in particular, the
effect of galactic outflows on the abundance of DLAs.  We employ the
``conservative entropy'' formulation of SPH developed by \citet{SH02a},
which mitigates against the systematic overcooling that affected 
earlier simulations.  In addition, we utilise a series of simulations 
of varying box-size and particle number to isolate the impact of 
numerical resolution on our results.

We show that the DLA abundance was overestimated in previous studies
for three reasons: (1) the overcooling of gas occurring with
non-conservative formulations of SPH, (2) a lack of numerical
resolution, and (3) an inadequate treatment of feedback.  Our new
results for the total neutral hydrogen mass density, DLA abundance,
and column density distribution function all agree reasonably well
with observational estimates at redshift $z=3$, indicating that DLAs
arise naturally from radiatively cooled gas in dark matter haloes that
form in a $\Lam$CDM universe.  Our simulations suggest a moderate
decrease in DLA abundance by roughly a factor of two from $z=4.5$ to
3, consistent with observations.  A significant decline in abundance
from $z=3$ to $z=1$, followed by weak evolution from $z=1$ to $z=0$, is
also indicated, but our low-redshift results need to be interpreted
with caution because they are based on coarser simulations than the
ones employed at high redshift.  Our highest resolution simulation
also suggests that the halo mass-scale below which DLAs do not exist is
slightly above $10^8\himsun$ at $z=3-4$, somewhat lower than
previously estimated.

\end{abstract}



\section{Introduction}
\label{section:intro}

Damped Lyman-$\alpha$ absorbers, historically defined as quasar
absorption systems with neutral hydrogen column density $\NHI>2\times
10^{20} \cm^{-2}$ \citep{Wol86}, are one of the best probes of structure
formation in the early universe.  Since DLAs are dense concentrations of
gas often found at $z\geq 3$, it is natural to suppose that they are
closely linked to the formation of galaxies and stars at high
redshift. It has become clear in recent years from the study of
Lyman-break galaxies at $z\sim 3-4$ \citep[e.g.][]{Ade98, Ste99,
Sha01} that the assembly of galaxies is actively going on at $z\sim
3$, consistent with hierarchical structure formation in a cold dark
matter universe
\citep[e.g.][]{Mo96, Bau98, Jin98, KHW99, Kau99, Mo99, Nag02, Wei02}.

A picture of the history of cosmic star formation emerging from both
theory and observation is that it rises from the present towards high
redshift, even beyond $z=3$ \citep[e.g.][]{Pas98, Bla99, Nag01a,
Lan02, SH02c, Her02}. The conversion of gas into stars is, therefore,
taking place at a significant rate at $z\geq 3$.  If DLAs dominate the
neutral hydrogen gas content of the Universe at $z\sim 3$, they are
thus serving as an important reservoir of neutral gas for star
formation.  Determining the physical nature and number density of DLAs
may hence be one of the most important keys for further constraining
the cosmic star formation history and theories of galaxy formation.

Although the current sample of observed DLAs at $z\geq 1.5$ is not yet
as large as that of Lyman-break galaxies (where $\approx 1000$ are
known), the total number of DLAs that have been discovered is now
approaching $\approx 100$, and the number density per unit redshift of
high column density systems appears to peak at around $z\sim 3$
\citep{Sto00}.  At lower redshift, the situation is rather different.
The identification of DLAs at $z<1.5$ has been difficult because they
are rare
and the need for ultra-violet (UV) spectroscopy
to detect Ly$\alpha$ absorption at these low redshifts.  The number of
quasars studied in UV has been small until recently. To overcome this
difficulty, \citet{Rao00} searched for DLAs in 87 Mg {\sc ii}
absorbers, and uncovered 12 new systems.  There are 23
DLAs at $z<1.65$ listed by \citet{Rao00}.

Despite the likely importance of DLAs and the accumulation of
observational data on them, their true nature remains controversial.
Historically, it has often been suggested that high-redshift DLAs are
large, rapidly rotating discs, because DLAs have properties similar to
local galactic discs, such as large neutral hydrogen column densities
together with low degrees of ionisation and small velocity dispersions
\citep{Wol86}.  More recently, \citet{Pro97, Pro98} argued that the
observed distribution of velocity widths and the asymmetric absorption
profiles of low-ionisation ionic species can be best described by
massive, rapidly rotating cold discs.

On the other hand, \citet{Hae98} examined a small number of dark
matter haloes in a high-resolution (sub-kpc) SPH simulation, and
showed that such observational signatures can also be explained by a
mixture of rotation, random motions, infall, and mergers of
protogalactic clumps.  There are some observational indications
\citep{LeB97, Rao98, Kul00, Kul01} from direct imaging studies that
luminous disc galaxies may not represent the dominant population of DLA
galaxies (i.e. galaxies that host DLAs).  Although the possibility of
artifacts due to point-spread-function effects cannot be fully
excluded, these observations suggest that some DLA galaxies could be
compact, clumpy objects, or low surface brightness galaxies,
rather than large, well-formed protogalactic disks or spheroids.

Robust numerical estimates of DLA properties have been hampered by the
significant requirements on numerical resolution needed to capture the
full population of DLAs.  Earlier studies by
\citet{Katz96-dla} and \citet{Her96} showed that the observed \HI
column density distribution can be reproduced within a factor of a few
in hydrodynamic simulations based on a CDM model over a wide range of
column densities $10^{14}\cm^{-2}\leq \NHI \leq
10^{22}\cm^{-2}$. Their results demonstrated that the Ly-$\alpha$
forest develops naturally in the hierarchical clustering scenario of
CDM universes, and that DLAs and Lyman-limit systems ($\NHI\geq
10^{17}\cm^{-2}$) arise in these models from radiatively cooled gas
inside dark matter haloes that host forming galaxies at high redshift.
However, their calculations were based on simulations of a
critical-density universe with $\Omega_{\rm m}=1$, and could not
resolve haloes with masses $M<10^{11}\himsun$.

Subsequently, \citet{Gar97a, Gar97b, Gar01} extended the earlier
results of \citet{Katz96-dla} and \citet{Her96} by developing a method
to correct for the resolution limitations of the simulations. They
measured the relation between absorption cross-section and halo
circular velocity from hydrodynamic simulations, and then convolved it
with the analytic halo mass function \citep[e.g.][]{Pre74, She99} to
compute the cumulative abundance of DLAs. Using this correction
method, they were able to reproduce the observed
abundance of DLAs {\it if} they required that haloes with circular
velocity $v_c \leq 60\kms$ (which corresponds to $M\approx 2\times
10^{10}\himsun$) did not harbour DLAs.  However, their
simulations could not resolve haloes with masses below $10^{10}\himsun$,
although such haloes may still host a significant number of DLAs.  If
the absorption cross-section of haloes with $M<10^{10}\himsun$ does
not follow the same relation between the cross-section and the halo
mass as determined from higher mass haloes, the DLA abundance could be
either over- or underestimated.  Simulations of higher resolution are
hence clearly needed to make more robust predictions for the DLA
abundance at redshift $z=3$.
 
Recently, \citet{SH02a} developed a novel formulation of SPH that is
based on integrating the entropy as an independent thermodynamic
variable \citep[e.g.][]{Lucy77, Hern93}, and which takes variations
of the SPH smoothing lengths self-consistently into account.  
They showed that this new version maintains contact discontinuities 
(as they arise at the interface between cold and hot gas in haloes) 
much better than previous treatments of SPH.  Consequently, their 
formulation does not suffer from the severe overcooling that was 
typically seen in previous SPH simulations.\footnote{The likelihood 
that earlier SPH studies were affected by overcooling due to numerical 
effects is supported by comparisons between our new formulation and 
simulations using an adaptive mesh refinement (AMR) algorithm 
(M. Norman, private communication).  This finding will be presented 
in due course.}  
We hence use this new methodology for our studies of DLAs.

\begin{table*}
\begin{center}
\begin{tabular}{cccccccc}
\hline
Run & Boxsize & ${N_{\rm p}}$ & $m_{\rm DM}$ & $m_{\rm gas}$ &
$\epsilon$ & $z_{\rm end}$ & wind \\
\hline
\hline
R3  & 3.375 & $2\times 144^3$ &  $9.29\times 10^5$ & $1.43\times 10^5$ &0.94 & 4.00 & strong \cr
R4  & 3.375 & $2\times 216^3$ &  $2.75\times 10^5$ & $4.24\times 10^4$ &0.63 & 4.00 & strong \cr
\hline
O3  & 10.00 & $2\times 144^3$ &  $2.42\times 10^7$ & $3.72\times 10^6$ &2.78 & 2.75 & none \cr
P3  & 10.00 & $2\times 144^3$ &  $2.42\times 10^7$ & $3.72\times 10^6$ &2.78 & 2.75 & weak \cr
Q3  & 10.00 & $2\times 144^3$ &  $2.42\times 10^7$ & $3.72\times 10^6$ &2.78 & 2.75 & strong \cr
Q4  & 10.00 & $2\times 216^3$ &  $7.16\times 10^6$ & $1.10\times 10^6$ &1.85 & 2.75 & strong \cr
Q5  & 10.00 & $2\times 324^3$ &  $2.12\times 10^6$ & $3.26\times 10^5$ &1.23 & 2.75 & strong \cr
\hline						                        
D4  & 33.75 & $2\times 216^3$ &  $2.75\times 10^8$ & $4.24\times 10^7$ &6.25 & 1.00 & strong \cr
D5  & 33.75 & $2\times 324^3$ &  $8.15\times 10^7$ & $1.26\times 10^7$ &4.17 & 1.00 & strong \cr
\hline						                        
G4  & 100.0 & $2\times 216^3$ &  $7.16\times 10^9$ & $1.10\times 10^9$ &12.0 & 0.00 & strong \cr
G5  & 100.0 & $2\times 324^3$ &  $2.12\times 10^9$ & $3.26\times 10^8$ &8.00 & 0.00 & strong \cr
\hline
\end{tabular}
\caption{Simulations employed in this study.
The box-size is given in units of $\himpc$, ${N_{\rm p}}$ is the
particle number of dark matter and gas (hence $\times\, 2$), $m_{\rm
DM}$ and $m_{\rm gas}$ are the masses of dark matter and gas
particles in units of $\himsun$, respectively, $\epsilon$ is the
comoving gravitational softening length in units of $\hikpc$,
and $z_{\rm end}$ is the ending redshift of the simulation. The
value of $\epsilon$ is a measure of spatial resolution.
From the top to the bottom row, we refer to R3 \& R4 collectively as
`R-series', the next 5 runs (O3 to Q5) are called `Q-series', D4 \& D5
are called `D-series', and G4 \& G5 are called `G-series'. The
`strong-wind' simulations form a subset of the runs analysed by
\citet{SH02c}.
\label{table:sim}}
\end{center}
\end{table*}

Our simulations also include a novel method for treating star
formation and feedback, as proposed by \citet{SH02b}. It is based on a
sub-resolution multi-phase description of the dense, star-forming
interstellar medium (ISM) and a phenomenological model for strong
feedback by galactic winds.  The inclusion of winds was motivated by
the possibility that outflows from galaxies at high redshift
\citep{Pet02} play a role in distributing metals into the
intergalactic medium \citep[e.g.][]{Aguirre01a, Aguirre01b}, and they
may also alter the distribution of neutral gas around galaxies
\citep{Ade02}, although this process remains uncertain
\citep[e.g.][]{Croft02, Kol03}.  Together with the increase in
numerical resolution provided by our simulations, it is of interest to
see how refinements in physical modelling modify the predictions
of DLA properties in a CDM universe.

In this paper, we focus on the abundance of DLAs in the redshift range
$z=0-4.5$.  The present work extends and complements earlier numerical
work by \citet{Katz96-dla} and \citet{Gar01}.  Physical properties of
DLAs such as their star formation rates, metallicities, and their
relation to galaxies will be presented elsewhere.

The paper is organised as follows. In
Section~\ref{section:simulation}, we briefly describe the numerical
parameters of our simulation set.  We then present the evolution of
the total neutral hydrogen mass density in the simulations in
Section~\ref{section:OmegaHI}.  In Section~\ref{section:cross}, we
describe how we compute the \HI column density and DLA cross-section
as a function of total halo mass.  In Section~\ref{section:abundance},
we determine the cumulative abundance of DLAs, and discuss the evolution
of DLA abundance from $z=4.5$ to $z=0$.  The \HI column density
distribution function is presented in Section~\ref{section:dist}. Finally,
we summarise and discuss the implication of our work in
Section~\ref{section:discussion}.


\section{Simulations}
\label{section:simulation}

We analyse a large set of cosmological SPH simulations that differ in
box size, mass resolution and feedback strength, as summarised in
Table \ref{table:sim}. In particular, we consider box sizes ranging
from 3.375 to $100\,h^{-1}{\rm Mpc}$ on a side, with particle numbers
between $2\times 144^3$ and $2\times 324^3$, allowing us to probe
gaseous mass resolutions in the range $4.2 \times 10^4$ to $1.1\times
10^9\,h^{-1}{\rm M}_\odot$.  These simulations are partly taken from
a study of the cosmic star formation history by \citet{SH02c},
supplemented by additional runs with weaker or no galactic winds. The
joint analysis of this series of simulations allows us to
significantly broaden the range of spatial and mass-scales that we can
probe compared to what is presently attainable within a single
simulation.

There are three main new features to our simulations.  First, we use a
new ``conservative entropy'' formulation of SPH \citep{SH02a} which
explicitly conserves entropy (in regions without shocks), as well as
momentum and energy, even when one allows for fully adaptive smoothing
lengths.  This formulation moderates the overcooling problem present
in earlier formulations of SPH \citep[see also][]{Yoshida02,
Pearce99, Croft01}.

Second, highly over-dense gas particles are treated with an effective
sub-resolution model for the ISM, as described by \citet{SH02b}.  In
this model, the dense ISM is pictured to be a two-phase fluid consisting
of cold clouds in pressure equilibrium with a hot ambient phase. Each
gas particle represents a statistical mixture of these phases. Cold
clouds grow by radiative cooling out of the hot medium, and this
material forms the reservoir of baryons available for star formation.
Once star formation occurs, the resulting supernova explosions deposit
energy into the hot gas, heating it, and also evaporate cold clouds,
transferring cold gas back into the ambient phase. This establishes a
tight self-regulation mechanism for star formation in the ISM.

Third, we implemented a phenomenological model for galactic winds in
order to study the effect of outflows on DLAs, galaxies, and the
intergalactic medium (IGM).  In this model, gas particles are stochastically
driven out of the dense star-forming medium by assigning an extra
momentum in random directions, with a rate and magnitude chosen to
reproduce mass-loads and wind speeds similar to those observed. See
\citet{SH02b} for a detailed discussion of this method.

Most of our simulations employ a ``strong'' wind of speed $484\,{\rm
km\,s^{-1}}$, but for the $10\,h^{-1}{\rm Mpc}$ box (runs O3, P3, Q3,
Q4, Q5; collectively called `Q-series') we also varied the wind
strength.  Therefore, this Q-series can be used to study both the
effect of numerical resolution and the consequences of feedback from
galactic winds.  The runs in the other simulation series then allow
the extension of the strong wind results to smaller scales
(`R-Series'), or to larger box-sizes and hence lower redshift (`D-'
and `G-Series').  Our naming convention is such that runs of the same
model (box-size and included physics) are designated with the same
letter, with an additional number specifying the particle resolution.

Our calculations include a uniform UV background radiation field of a
modified \citet{Haa96} spectrum, where reionisation takes place at
$z\simeq 6$ \citep[see][]{Dave99}, as suggested by observations
\citep[e.g.][]{Beck01} and radiative transfer calculations of the
impact of the stellar sources in our simulations on the IGM
\citep[e.g.][]{Sok03}.  The radiative cooling and heating rate is
computed as described by \citet{Katz96} assuming that the gas is 
optically thin and in ionization equilibrium.
The abundance of different ionic species, including H$^0$, He$^0$, H$^+$, 
He$^+$, and He$^{++}$, is computed by solving the network of equilibrium 
equations self-consistently with a specified UV background radiation.
The results presented in this paper should not be affected by the
assumption of ionization equilibrium as we are dealing with high 
density regions where this assumption is satisfied well.  
The adopted cosmological parameters of all runs are 
$(\Om,\Ol,\Ob,\sigma_8, h)= (0.3, 0.7, 0.04, 0.9, 0.7)$. 
The simulations were performed on the Athlon-MP
cluster at the Center for Parallel Astrophysical Computing (CPAC) at
the Harvard-Smithsonian Center for Astrophysics, using a modified
version of the parallel {\small GADGET} code \citep{Gadget}.


\section{Neutral Hydrogen Mass Density}
\label{section:OmegaHI}

\begin{figure*}
\begin{center}
\resizebox{8.2cm}{!}{\includegraphics{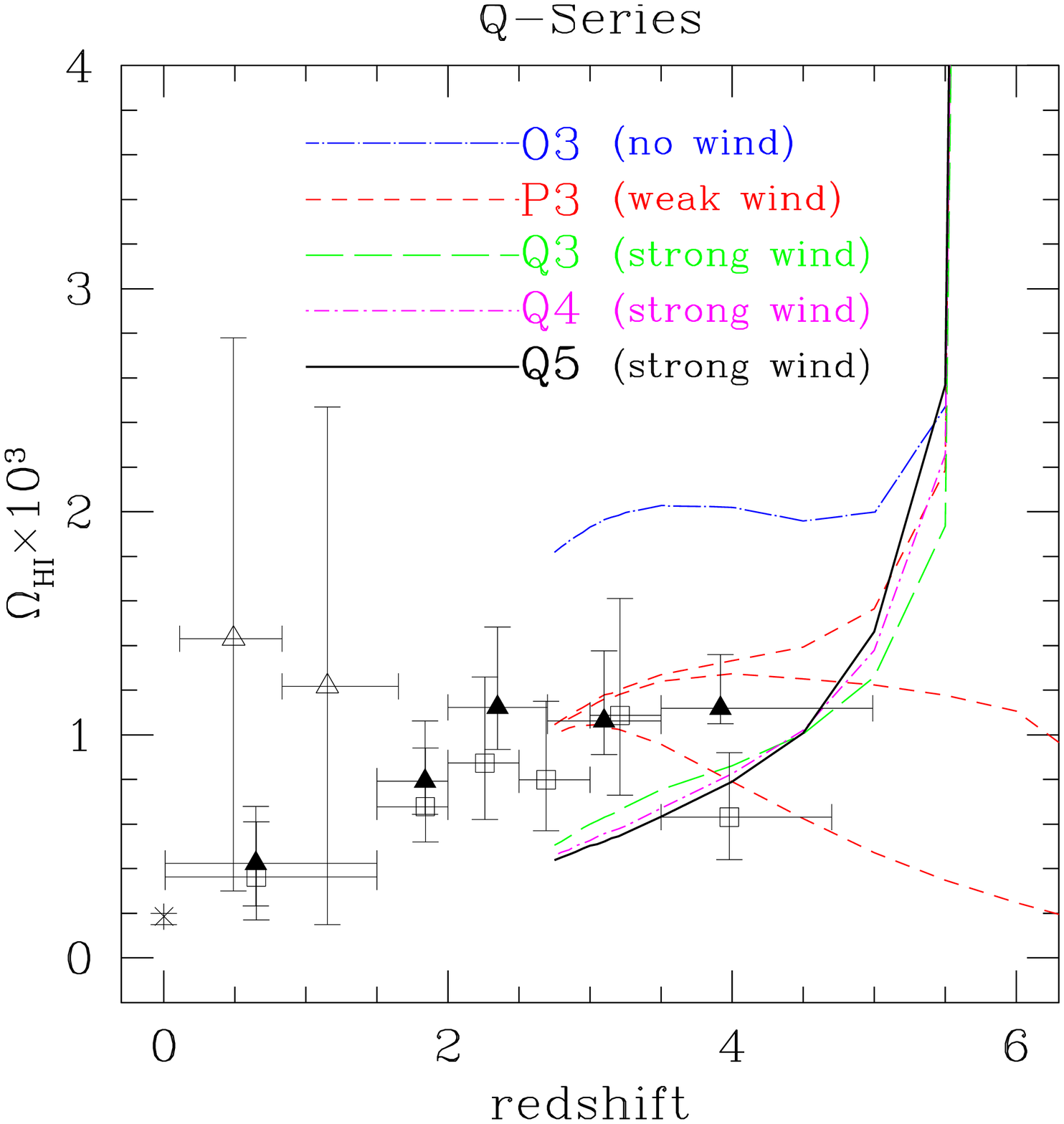}}%
\hspace{0.3cm}
\resizebox{8.2cm}{!}{\includegraphics{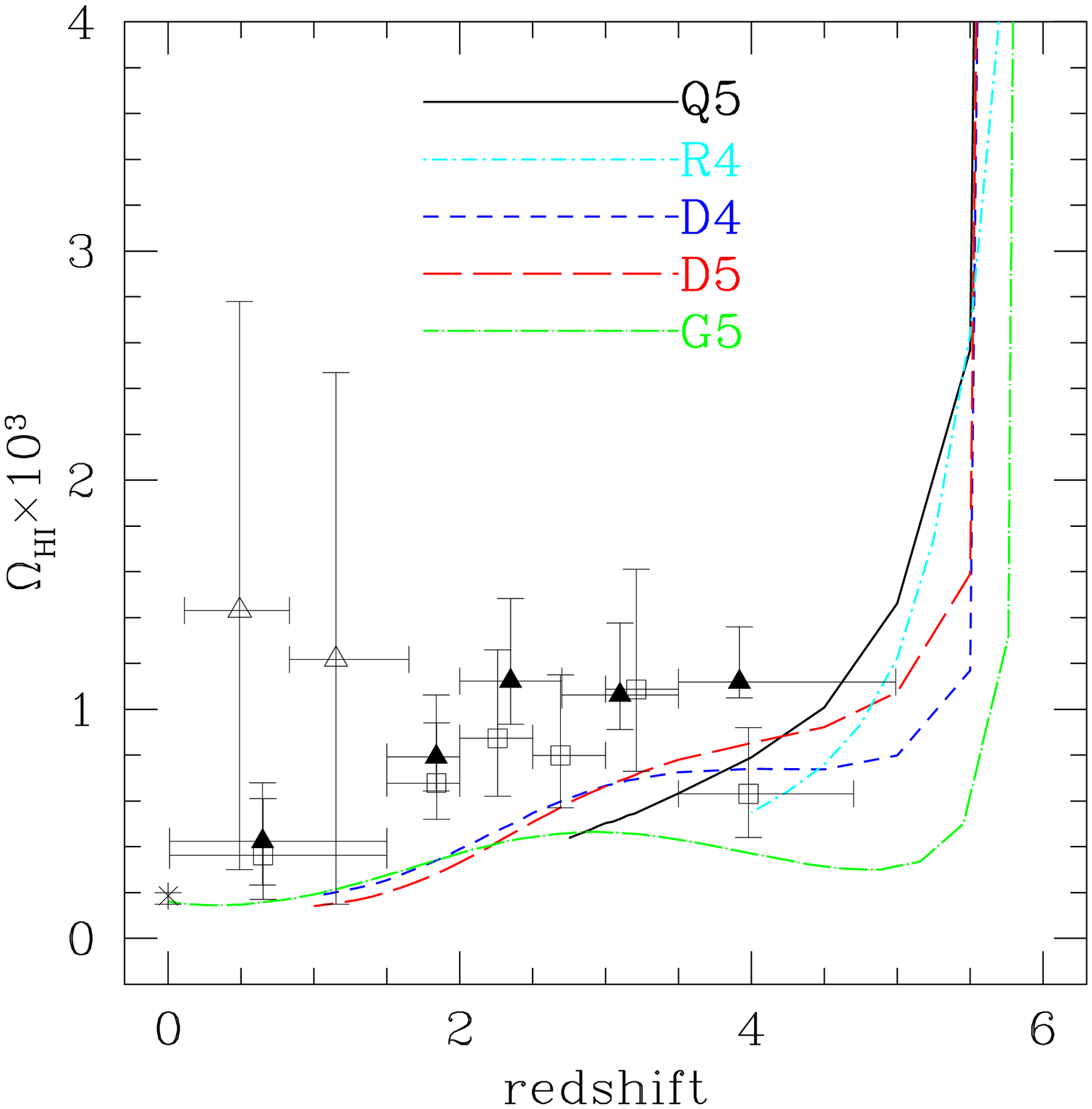}}\\%
\caption{Evolution of the total neutral hydrogen mass density in each
simulation box as a function of redshift.  The plotted values are
$\OHI\times 10^3$.  We also show observational data points from
\citet[][open squares; only for DLAs]{Sto00}, \citet[][filled
triangles; including the correction for the neutral gas not included
in DLAs]{Per01}, \citet[][open triangles]{Rao00}, and \citet[][open
cross at $z=0$]{Zwa97}.  {\it Left panel:} A comparison of the
Q-series (runs in $10\,h^{-1}{\rm Mpc}$ boxes) is shown. The decrease
in $\OHI$ from O3 (no wind) to P3 (weak wind), and then to Q3 (strong
wind), shows the effect of feedback by galactic winds.  The comparison
between Q3, Q4, and Q5 shows the level of convergence achieved for
runs with different resolution.  For P3 (top short-dashed line), we
separately show $\OHI$ in regions of overdensity $1+\delta>10^3$ and
$10^4$ (middle and bottom short-dashed line, respectively).  
{\it Right panel:} Results for the R-, D-, and G-series are
shown. Q5 is also included to ease comparison with the left
panel. Results for R3 and G4 are omitted for clarity (see text).
\label{omega.eps}}
\end{center}
\end{figure*}

In our simulation methodology, gas is subject to a thermal instability
yielding a multiphase medium once the physical gas density lies
above a threshold density $\rho_{\rm th}$, which marks the onset of
cold cloud formation. If the physical density is lower than this
threshold, a particle represents ordinary
gas in a single phase.  In this latter case, the neutral hydrogen mass
of the particle can be computed as follows:
\beq m_{\rm HI} = N_{\rm h} \cdot
\XH \cdot m_{\rm gas} \quad (\rho<\rho_{\rm th}), \eeq
\noindent
where $\XH=0.76$ is the primordial mass fraction of hydrogen, and
$N_{\rm h}$ is the number density of neutral hydrogen atoms in units
of the total number density of hydrogen nuclei. The quantity $N_{\rm
h}$ is computed by solving the ionisation balance as a function of 
density and current UV background flux.

If the gas density is higher than the threshold density, we identify
the mass of neutral gas with the mass of cold clouds contained in the
multiphase medium. The mass of neutral hydrogen of such a multiphase
particle is then given by
\beq
m_{\rm HI} = x \cdot \XH \cdot m_{\rm gas} \quad (\rho>\rho_{\rm th}),
\eeq
where $x$ is the mass fraction of cold clouds.  In the multiphase
model of Springel \& Hernquist (2002b), $x$ can be computed as 
\beq
x\equiv \rho_c/\rho =
1+\frac{1}{2y}-\sqrt{\frac{1}{y}+\frac{1}{4y^2}}, 
\label{eqxx1}
\eeq 
where the quantities $y$ and $\rho_{\rm th}$ are defined by 
\beq y \equiv
\frac{t_* \Lambda_{\rm net}(\rho,u_h)}{\rho[\beta u_{\rm
SN}-(1-\beta)u_c]} \label{eqxx2}
\eeq and 
\beq \rho_{\rm th} =
\frac{x_{\rm th}}{(1-x_{\rm th})^2} \frac{\beta u_{\rm SN}-(1-\beta)u_c}{t_*^0
\Lambda(u_{\rm SN}/A_0)}.  
\label{eqxx3}
\eeq 
Here, $\Lambda_{\rm net}(\rho,u)$ is the usual cooling function, and we
have defined $\Lambda(\rho,u)\equiv \Lambda_{\rm net}(\rho,u)/\rho^2$.
The parameter $\beta$ gives the mass fraction of short-lived stars
that explode as supernovae, $u_{\rm SN}$ describes the energy released
by the supernovae, $u_c$ is the assumed temperature of the cold
clouds, $u_h$ the temperature of the hot medium, $A_0$ the cloud
evaporation parameter, and $t_*^0$ gives the star formation time-scale.
The quantity 
$x_{\rm th} = (u_h-u_4)/(u_h-u_c) \simeq 1-A_0u_4/u_{\rm SN}$ is the
mass fraction in cold clouds at the threshold density, where $u_4$ is
the specific energy corresponding to $T=10^4~{\rm K}$. We refer to
Springel \& Hernquist (2002b) for a more detailed explanation of these
parameters, and a derivation of equations (\ref{eqxx1}) to
(\ref{eqxx3}).

In Figure \ref{omega.eps}, we show the total neutral hydrogen mass
density as a function of redshift.  The values plotted are given in
terms of $\OHI\times 10^3$.  At redshifts above six, essentially all the
hydrogen in the simulation box is still neutral ($\OHI \simeq \XH \Ob
= 0.0304$).  Once the ionising background sets in at $z=6$, the
neutral hydrogen starts to become ionised and the neutral fraction
decreases rapidly by a few orders of magnitude; i.e.~reionisation
takes place. Note that with the exception of the R-Series, our
earliest simulation output at $z<6$ corresponds to $z=5.5$. This is
why most of the models in Figure~\ref{omega.eps} seem to show a rapid
rise of $\OHI$ already at $z=5.5$, which simply arises by drawing a
line to the data point at $z=6$ which lies a few orders of magnitude
higher in this linear plot.

We also include observational data points, for comparison. The data
points from \citet[][open squares]{Sto00} account only for DLAs, but
those of \citet[][filled triangles]{Per01} include a correction for
the neutral gas that is in sub-DLAs $(10^{19}<\NHI<10^{20}\cm^{-2}$).
Data points from \citet[][solid triangles]{Rao00} at low-redshift and
\citet[][open cross]{Zwa97} at $z=0$ are also shown.

In the left panel of Figure~\ref{omega.eps}, we compare results only
for the Q-series ($10\,h^{-1}{\rm Mpc}$ box), allowing us to
assess convergence as a function of mass resolution and to investigate
the dependence of the results on the strength of feedback from winds.

The comparison between Q3, Q4, and Q5 (strong wind) shows that there
is quite good agreement between runs with different numerical
resolution. In fact, the results for Q3, Q4, and Q5 are essentially
identical at $z\simeq 4.5$, with Q5 being slightly higher for $z\geq
4.5$ than its lower resolution counterparts, while for $z\leq 4.5$ the
opposite trend is observed.  This mild effect can be understood
as follows: in a higher resolution run like Q5, many more small dark
matter haloes can be resolved than in a lower resolution run (like
Q3), particularly at early times, where gas can cool very efficiently.
As a result, the higher resolution simulation develops a larger
fraction of cold and hence neutral gas at high redshift.  However, an
increased fraction of cold gas will also trigger more intense
star formation that both consumes neutral gas and leads to gas
ejection by winds from low mass haloes. Subsequently, the neutral
fraction can then fall slightly below the lower resolution runs.

Comparing the results for the models O3 (no wind), P3 (weak wind), and
Q3 (strong wind) shows the impact of feedback by galactic winds.  As
the wind strength increases, the neutral density $\OHI$ decreases.
More neutral gas is then ejected out of the dense ISM into the
intergalactic medium, where it becomes highly ionised by UV
background radiation.  Interestingly, `O3' (no wind run) exceeds
all observed data points, so a feedback effect such as galactic winds
appears necessary to make the $\OHI$ measurements of the simulations
consistent with observations.  The results for our `strong-wind' runs
(Q3, Q4, Q5) underpredict the observational estimates at $z=3$
slightly, but there is still marginal agreement within $1\,\sigma$,
which is encouraging. However, the best value for the galactic wind
strength parameter for our simulation seems to lie somewhere between
that of P3 (weak wind) and the Q-runs (strong wind).

For the `P3' run, we also show separate measurements of $\OHI$
restricted to regions of overdensity $1+\delta >10^3$ and $10^4$,
respectively (red short-dashed lines). The fact that the lines for
$1+\delta>10^4$ and $10^3$ have converged by $z\simeq 3$ shows that
most of the neutral hydrogen mass in the universe is already in a
highly concentrated form by this epoch.

In the right panel of Figure~\ref{omega.eps}, we show our results for
simulations of the R-, D-, and G-series, together with Q5 for
reference to the left panel.  The results for D4 and D5 are consistent
with one another at $z=3$.  `R3' is not shown because it is almost
identical to `R4', and `G4' is omitted because it underpredicts $\OHI$
significantly due to lack of resolution at $z\geq 3$.
By comparing to the simulations of the Q- and D-series, we see that
the resolution of the G-series is not sufficient to correctly 
describe
the neutral fraction at $z=3$.  This is because even the $2\times
324^3$ run G5 misses the neutral gas content in large numbers of small
dark matter haloes that are present in the higher resolution runs at
$z=3$, such as those of the Q-series.  Therefore, we consider Q5 to be
the most reliable run at $z=3$ among our simulation set.  We also see
that $\OHI$ of `R4' is lower than that of `Q5', despite the fact
that the R-series has higher mass resolution than the Q-series.  This
is likely due to the rather small box-size of the R-series compared to
the Q-series, which leads to an insufficient sampling of rare, massive
objects, and compromises the use of R4 as a truly representative
sample of the universe.

The effect of the multiphase model adopted in the current simulations 
can be assessed by setting the value of cold gas mass fraction to $x=1$
for the multiphase gas particles [see Equation (\ref{eqxx1})]. We find
that the value of $\OHI$ becomes larger by about 15\% in such a case.
This suggests that previous formulations of hydrodynamic simulations 
without a consideration for the multiphase nature of the gas would have 
overestimated the cold gas fraction by a similar amount.


\section{\HI Column Density \& DLA Cross-Section}
\label{section:cross}

We now describe how we compute the \HI column density $\NHI$ and the
DLA cross-section $\sdla$ for each dark matter halo.  First, we
identify dark matter haloes by applying a conventional
friends-of-friends algorithm to the dark matter particles in each
simulation. We set the minimum number of dark matter particles for a
halo to 32; i.e.~haloes with fewer particles are not included in the
group catalogue.  We have confirmed that the dark matter halo mass
functions agree well with the analytic mass function of \citet{She99}.
After dark matter haloes are identified, we associate each gas and
star particle with their nearest dark matter particle, including them
in the particle list of the corresponding haloes, when appropriate.

Then, for each halo, a uniform grid covering the entire halo and
whose grid-size is equal to the gravitational softening length, is
placed at the center-of-mass of the halo. We then project the neutral
gas in the halo onto a plane, and obtain the column
density of each grid-cell in this plane.  The neutral mass of each
gas particle is smoothed over a spherical region of grid-cells,
weighted by the SPH kernel.  To check the robustness of the result, we
also tried a cloud-in-cell assignment scheme where the neutral mass
of each gas particle is uniformly distributed over a cubic region of
size $\ell=(\frac{4\pi}{3})^{1/3}s$ and
$\frac{1}{2}(\frac{4\pi}{3})^{1/3}s$ centered on the particle.  Here
$s$ is the SPH smoothing length. Differences in the smoothing method
can lead to slight differences in the \HI column density distribution,
as we will discuss later in Figure~\ref{dist_z3.eps}. In the
following, we adopt the SPH smoothing method for our primary results
unless explicitly stated otherwise.

Once the comoving neutral mass density $\rho_{i,\rm HI}$ in each
grid-cell of volume $\epsilon^3$ is known, it is straightforward to
project the density distribution along the direction perpendicular to
the plane to obtain the column density as 
\beq 
\NHI = \sum_i \rho_{i,{{\rm HI}}} \cdot \epsilon / m_p \cdot (1+z)^2, 
\eeq 
where $\epsilon$ is the comoving gravitational softening length, 
$m_p$ is the proton mass, and $z$ is the redshift.

Note that in the present study, we do not apply a self-shielding
correction when computing the neutral hydrogen fraction. As
\citet{Katz96-dla} have shown, damped systems with column densities
above $\NHI\simeq 10^{20}\cm^{-2}$ are essentially fully neutral and
are not affected by self-shielding.  The correction is
expected to be large for systems with $10^{17}\leq \NHI \leq
10^{20}\cm^{-2}$, however.  A full 3-dimensional treatment of
self-shielding is beyond the scope of the present study, but it is
clearly an interesting and important issue in its own right. The
results presented in this paper for very high column density systems
should however be robust against self-shielding corrections.

Once the column density of each cell in the projected plane is obtained,
we estimate the comoving DLA cross-section of each halo by simply
counting the number of grid-cells that exceed $\NHI = 2\times
10^{20}\cm^{-2}$ and multiplying this number by the comoving unit area
$\epsilon^2$ of the grid-cells.


\subsection{DLA cross-section at redshift 3}
\label{section:z3_area}

\begin{figure*}
\epsfig{file=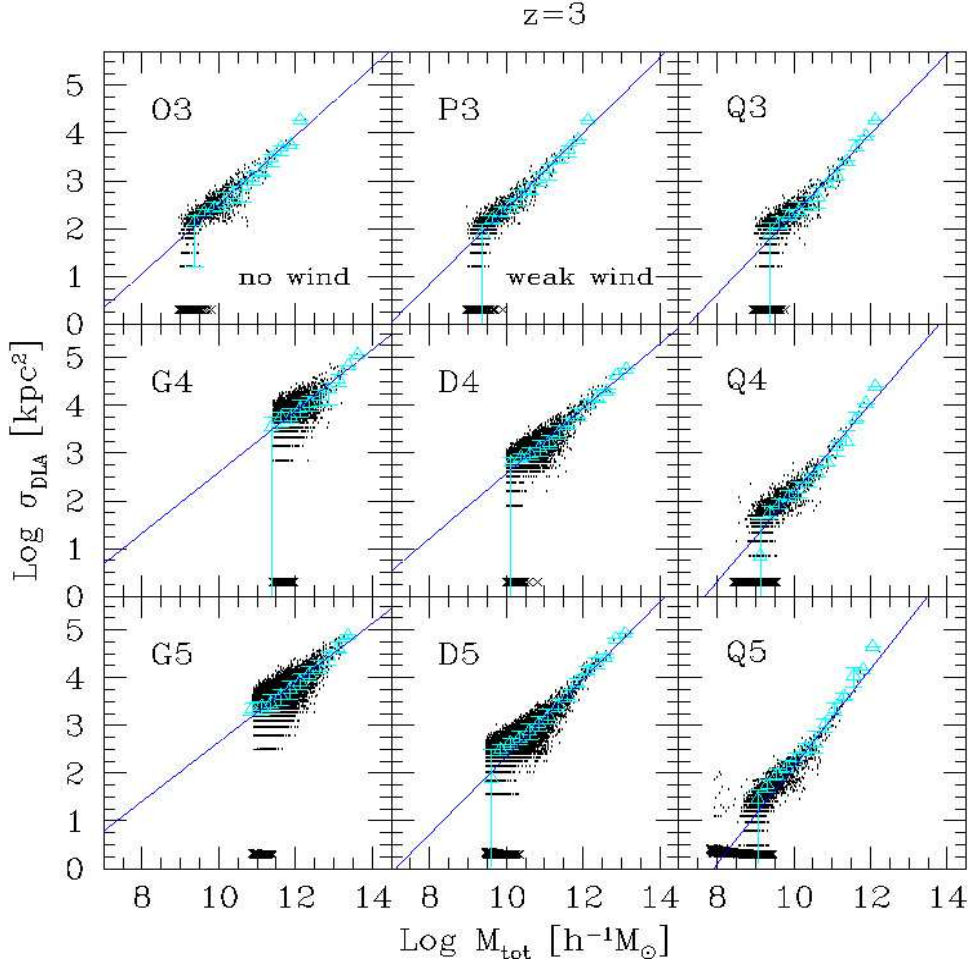,height=5in,width=5in, angle=0} 
\caption{Comoving DLA cross-section $\sdla$ as a function of total
halo mass at $z=3$ for the Q-, D-, and G-series. All panels are for runs
with `strong winds', except where labelled otherwise.  We show
median values for logarithmic mass bins as open triangles, and the
quartiles in each bin are indicated in the form of error bars.  The
horizontal concentration of crosses at $\log\sdla=0.3$ indicates 
haloes without DLAs.  The solid lines are power-law fits to the
median points as described in the text.
\label{area_z3.eps}}
\end{figure*}

In Figure~\ref{area_z3.eps}, we show the comoving DLA cross-section
$\sdla$ as a function of total halo mass ${M_{\rm tot}}$ at redshift
$z=3$.  All panels are for runs that include `strong winds', except
where explicitly labeled otherwise.  The data points are binned in
terms of $\log\Mtot$, and the median value in each bin is shown by the
open triangles.  The quartiles in each mass bin are shown as error
bars.  For plotting purposes only, we assign an arbitrary value 
of $\log\sdla=0.3$ to all haloes with no DLAs, and they are shown by 
the crosses at the bottom of each
panel.  We included these `no-DLA haloes' in computing the median
cross-section, therefore the error bars in the lowest mass-bins
sometimes extend to the bottom of the figure.

We then fit the median points to a power law, $\sdla \propto \Mtot
^\alpha$, assuming a functional form of
\beq
\label{eq:powerlaw}
\log\sdla = \alpha (\log\Mtot-12)+\beta, \eeq 
and determine the
values of the slope `$\alpha$' and the normalisation `$\beta$' by
least-squares fitting.  The value of $\beta$ hence gives the value of
$\log\sdla$ at $\Mtot=10^{12}\himsun$.  We chose this reference
mass-scale because it is well covered by most of the simulations used
in this paper.

Unlike the analysis of \citet{Gar97a}, we do not invoke a limiting
halo mass below which a dark matter halo does not harbour a DLA.  As
can be seen in all the panels of Figure~\ref{area_z3.eps}, such a
clear cutoff does not really exist, and DLAs continue to 
be found in
haloes with masses down to $\Mtot \simeq 10^{8.3}\himsun$ in `Q5'.  
We will come back to this point later.

\begin{table}
\begin{center}
{\sc Redshift 3} \\
\begin{tabular}{cccc}
\hline
Run & slope $\alpha$ & $\beta$ \cr
\hline
\hline
O3 & 0.72 & 3.94 \\
P3 & 0.79 & 3.99 \\
Q3 & 0.84 & 3.98 \\
Q4 & 0.93 & 4.03 \\
Q5 & 1.02 & 4.18 \\
\hline
D4 & 0.68 & 3.93 \\
D5 & 0.81 & 3.96 \\
\hline
G4 & 0.64 & 3.88 \\
G5 & 0.62 & 3.89 \\
\hline
\end{tabular}
\caption{The parameters obtained by least-square fitting the power-law
of equation~(\ref{eq:powerlaw}) to the median points shown in
Figure~\ref{area_z3.eps}.
\label{table:z3fit}}
\end{center}
\end{table}

We summarise the results of our power-law fitting in
Table~\ref{table:z3fit}.  It is satisfying that the values of the
normalisation `$\beta$' agree very well among different runs. This
demonstrates that our results for runs with widely varying resolution
are numerically well-converged at the mass-scale of
$\Mtot=10^{12}\himsun$.  It is seen that the slope `$\alpha$' becomes
steeper as the galactic wind strength increases from O3 to P3, and
further to Q3.  This is because gas in low-mass haloes is lost at a
higher rate in runs with stronger winds, making the DLA cross-sections
decrease for small haloes.  Another trend seen in
Table~\ref{table:z3fit} is that the slope becomes somewhat steeper as
the resolution of the simulation increases from Q3 to Q4, and then to
Q5.  This can be explained by the fact that higher resolution
simulations can resolve star formation in small haloes, leading to
ejection of gas out of them, lowering their content of neutral gas. On
the other hand, a lower resolution simulation misses this star
formation, resulting in an overestimate of the baryon and neutral
fraction in the first generation of haloes that is `seen' in the
simulation.

\citet{Gar01} reported a slightly shallower slope even compared to
`O3' (no-wind run): $\sdla\propto v_c^{1.57} \propto M^{0.52}$ (see
Table 2 of their paper).  Here, $v_c$ is the circular velocity of a
halo, related to the halo mass by $v_c \propto {M}^{1/3}$. A shallower
slope in general implies a higher abundance of DLAs.  A number of
effects are responsible for this difference: (1) Their resolution was
slightly lower (${N}=128^3$) than that of our O3/P3/Q3-runs. (2) The
overcooling problem in previous formulations of SPH may have caused
the slope to be shallower owing to an overestimate of
the neutral gas fraction,
particularly in small haloes.  With our new `conservative entropy' 
formulation of
SPH, the cold gas fraction in haloes is expected to be lower, although
the magnitude of this effect as a function of halo mass is not fully
clear.  (3) Their treatment of feedback is known to be inefficient,
because thermal energy injected into the gas is radiated away 
very rapidly.  However, given that our `O3'-run overpredicts $\OHI$
(Figure~\ref{omega.eps}) at $z=3$, some form of strong feedback seems
necessary to provide agreement with the observations.  
Noting that the slope of the power-law fit
steepens as the wind strength and resolution increase, we hence
conclude that the slope of \citet{Gar01} was probably too shallow.
This conclusion will be strengthened when we discuss the abundance of
DLAs in Section~\ref{section:abundance} and the column-density
distribution function in Section~\ref{section:dist}.


\subsection{DLA cross-section at redshift 4.5}
\label{section:highz_area}

\begin{figure*}
\epsfig{file=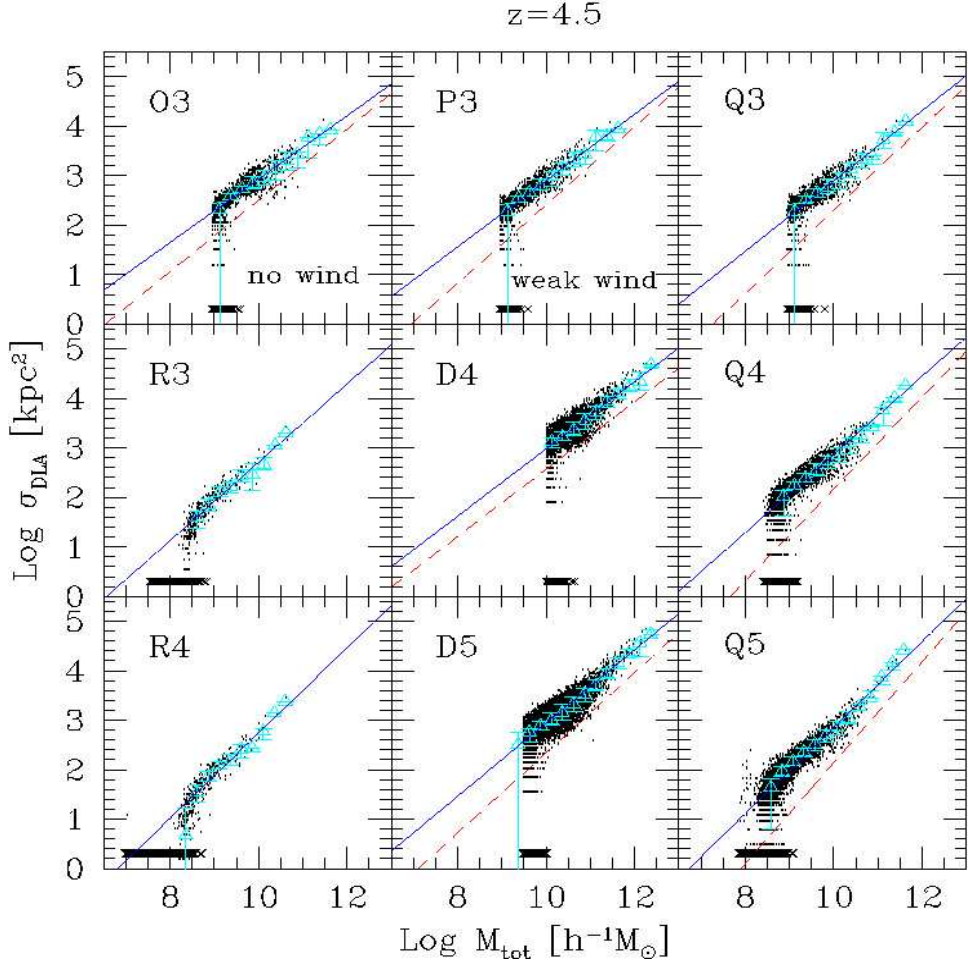,height=5in,width=5in, angle=0} 
\caption{Same as Figure~\ref{area_z3.eps}, but for $z=4.5$.  Here,
instead of G4 and G5, we show results for R3 and R4. The solid slanted
lines are the best-fits to the median points (indicated by the open
triangles) with a power-law (equation~\ref{eq:powerlaw}), and the
short-dashed lines are the fits we obtained at $z=3$, for comparison.
The horizontal concentration of crosses at $\log\sdla=0.3$ indicates haloes
that do not harbour a DLA.  Note that the range 
along the x-axis is different from 
that of Figure~\ref{area_z3.eps}.
\label{area_z4.5.eps}}
\end{figure*}

\begin{table}
\begin{center}
{\sc Redshift 4.5} \\
\begin{tabular}{cccc}
\hline
Run & slope $\alpha$ & $\beta$ \cr
\hline
\hline
R3 & 0.79 & 4.29 \\
R4 & 0.86 & 4.46 \\
\hline
O3 & 0.64 & 4.21 \\
P3 & 0.67 & 4.24 \\
Q3 & 0.71 & 4.31 \\
Q4 & 0.79 & 4.44 \\
Q5 & 0.87 & 4.59 \\
\hline
D4 & 0.68 & 4.34 \\
D5 & 0.74 & 4.42 \\
\hline
G4 & 0.56 & 4.31 \\
G5 & 0.65 & 4.30 \\
\hline
\end{tabular}
\caption{
The parameters obtained by least-squares fitting the power-law
of equation~(\ref{eq:powerlaw}) to the median points shown in
Figure~\ref{area_z4.5.eps}.
\label{table:z4.5fit}}
\end{center}
\end{table}

In Figure~\ref{area_z4.5.eps}, we show the DLA cross-section at
$z=4.5$ as a function of total halo mass. As before, the solid lines
show power-law fits that were obtained as described in the previous
subsection, while the short-dashed lines are the fits at redshift
$z=3$, for comparison.  Note that we do not plot results for the
G-series but instead show the R-series, which has much higher resolution,
but was evolved only to $z=4$ due to its small box-size.  The
results of the power-law fitting are summarised in
Table~\ref{table:z4.5fit}.  Similar to $z=3$, the values of the
normalisation `$\beta$' agree very well with each other between
runs of differing resolution. The generic trends in the slope as a
function of wind strength and resolution are also similar to what we
saw for $z=3$.

At the low-mass end of runs
R3 and R4, we see that the DLA cross-section is
dropping prominently at $\Mtot\sim 10^{8.3}\himsun$, strongly
departing from the fitted power-law.  To avoid being affected by 
this turn-down, we do not use the median points below 
$\log\sigma_{\rm DLA}=1.0$ for our power-law fitting.  A similar feature 
is seen in Q4 and Q5 at the same mass-scale.  Note that the resolution 
of the R-series is sufficient to resolve haloes with $\Mtot=10^8\himsun$
quite well, so this downturn in DLA cross-section is unlikely to be a
resolution artifact.  Instead, it is probably due to the physical effect
that the gas in these haloes is easily photoevaporated by the ionising
background, and/or ejected from haloes due to supernovae feedback.
At $z=3$, the mass-scale of $10^8 - 10^{8.5}\himsun$ corresponds to a 
circular velocity of $10-15\kms$, or a virial temperature slightly 
below $10^4\,{\rm K}$.  Note that this is a smaller mass-scale 
than was suggested by \citet{Qui96} and \citet{Tho96}, who
argued that haloes with circular velocities less than $40\kms$ are 
unlikely to harbour DLAs.  We will discuss this point further in 
Section~\ref{section:discussion}.


\subsection{DLA cross-section at lower redshift}
\label{section:lowz_area}

\begin{figure}
\epsfig{file=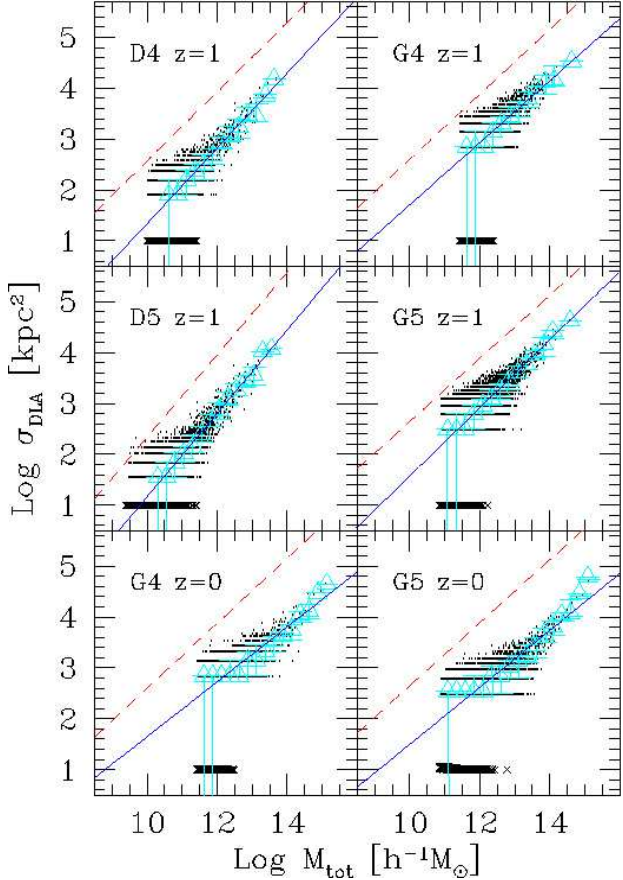,height=4.5in,width=3.2in, angle=0} 
\caption{Same as Figure~\ref{area_z4.5.eps}, but for $z=1$ and $0$.
The short-dashed lines are the fits we obtained at $z=3$, for
comparison.
\label{area_lowz.eps}}
\end{figure}

In Figure~\ref{area_lowz.eps}, we show DLA cross-sections as a function of
total halo mass for $z=1$ and $z=0$, and the parameters of
the fitted power-laws are summarised in Table~\ref{table:lowzfit}.  A
similar trend in the slope as a function of resolution exists at $z=1$
as we saw at $z=3$.  It is clear that the slope cannot be determined 
reliably for G4 and G5 at $z=0$ (and possibly at $z=1$) due to limited 
resolution, as is evident from the `stripes' seen at low cross-sections 
in the bottom two panels of Figure~\ref{area_lowz.eps}.

\begin{table}
\begin{center}
{\sc Lower Redshifts} \\
\begin{tabular}{cccc}
\hline
Run & slope $\alpha$ & $\beta$ \cr
\hline
\hline
z=1\\
\hline
D4 & 0.74 & 2.82 \\
D5 & 0.82 & 2.75 \\
\hline
G4 & 0.61 & 2.93 \\
G5 & 0.67 & 2.89 \\
\hline
\hline
z=0 \\
\hline
G4 & 0.54 & 2.73 \\
G5 & 0.56 & 2.62 \\
\hline
\end{tabular}
\caption{
The parameters obtained by least-square fitting the power-law
of equation~(\ref{eq:powerlaw}) to the median points shown in
Figure~\ref{area_lowz.eps}.
\label{table:lowzfit}}
\end{center}
\end{table}


\section{Cumulative Abundance of DLAs}
\label{section:abundance}

Dark matter haloes with masses below the resolution limit of a
simulation cannot be resolved. This is a serious problem when one
tries to compute the number density of DLAs based on a cosmological
simulation that does not resolve all small mass haloes that may host a
DLA.  Note in particular that the number density of dark matter haloes
is known to increase strongly towards lower masses. Even a small
incompleteness at low masses will hence prevent a reliable estimate of
the DLA abundance if only a simple number count of DLAs found in a
cosmological simulation is used.  

To overcome this limitation, \citet{Gar97a,Gar97b,Gar01}
convolved a theoretical fit to the dark matter halo mass function with
the measured relationship between DLA cross-section and halo mass.  In
this way, they were able to correct for incompleteness in the resolved
halo abundance of the simulations.  The cumulative abundance 
(or equivalently the rate of incidence) of DLAs per unit redshift as 
a function of halo mass in this approach can be expressed as 
\beq
\label{eq:abundance}
\frac{{\rm d}N_{\rm DLA}}{{\rm d}z}(>M, z) = \frac{{\rm d}r}{{\rm d}z} 
\int_M^{\infty} n_{\rm dm}(M',z)~\sdla(M',z)\,{\rm d}M',
\eeq
where $n_{\rm dm}(M,z)$ is the dark matter halo mass function (for
which we use the \citet{She99} parameterisation), and ${\rm d}r/{\rm
d}z = c/H(z)$ with $H(z)=H_0 E(z) = H_0\sqrt{\Om(1+z)^3+\Ol}$ for a
flat universe.  In order to carry out this integral, the power-law
fits obtained in Section~\ref{section:cross} can be used to represent
$\sdla(M,z)$ which give the mean relation between the halo mass and
the DLA cross-section.  Note that the dependence on the Hubble constant
disappears on the right-hand-side 
of equation (8) because ${\rm d}r/{\rm d}z$ scales as 
$\hinv$, while $n_{\rm dm}{\rm d}M$ depends on $h^3$, and $\sdla$ scales 
as $h^{-2}$ in the simulation.


\subsection{DLA abundance at redshift 3}
\label{section:abundance_z3}

\begin{figure*}
\begin{center}
\resizebox{8.2cm}{!}{\includegraphics{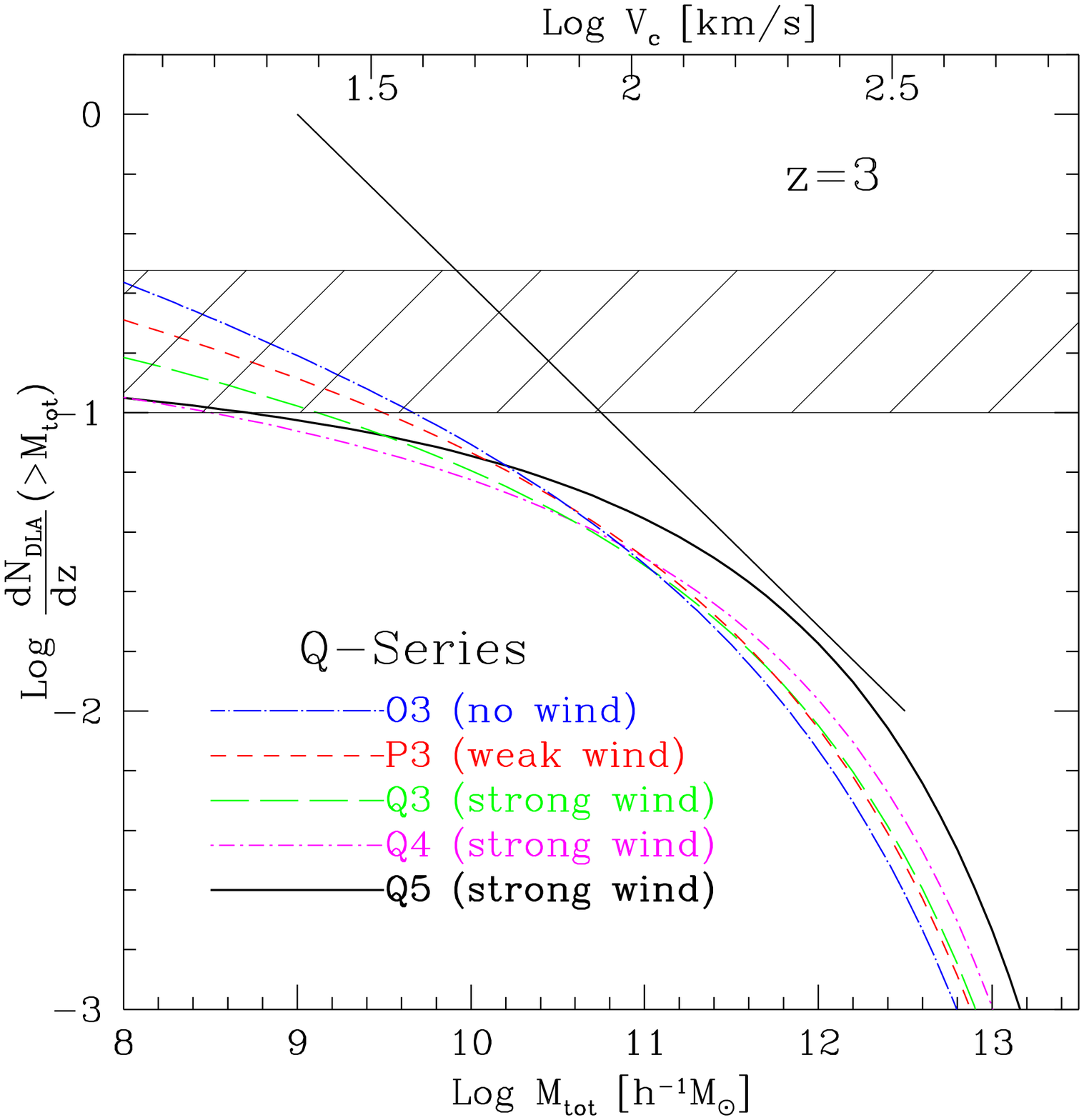}}%
\hspace{0.3cm}
\resizebox{8.2cm}{!}{\includegraphics{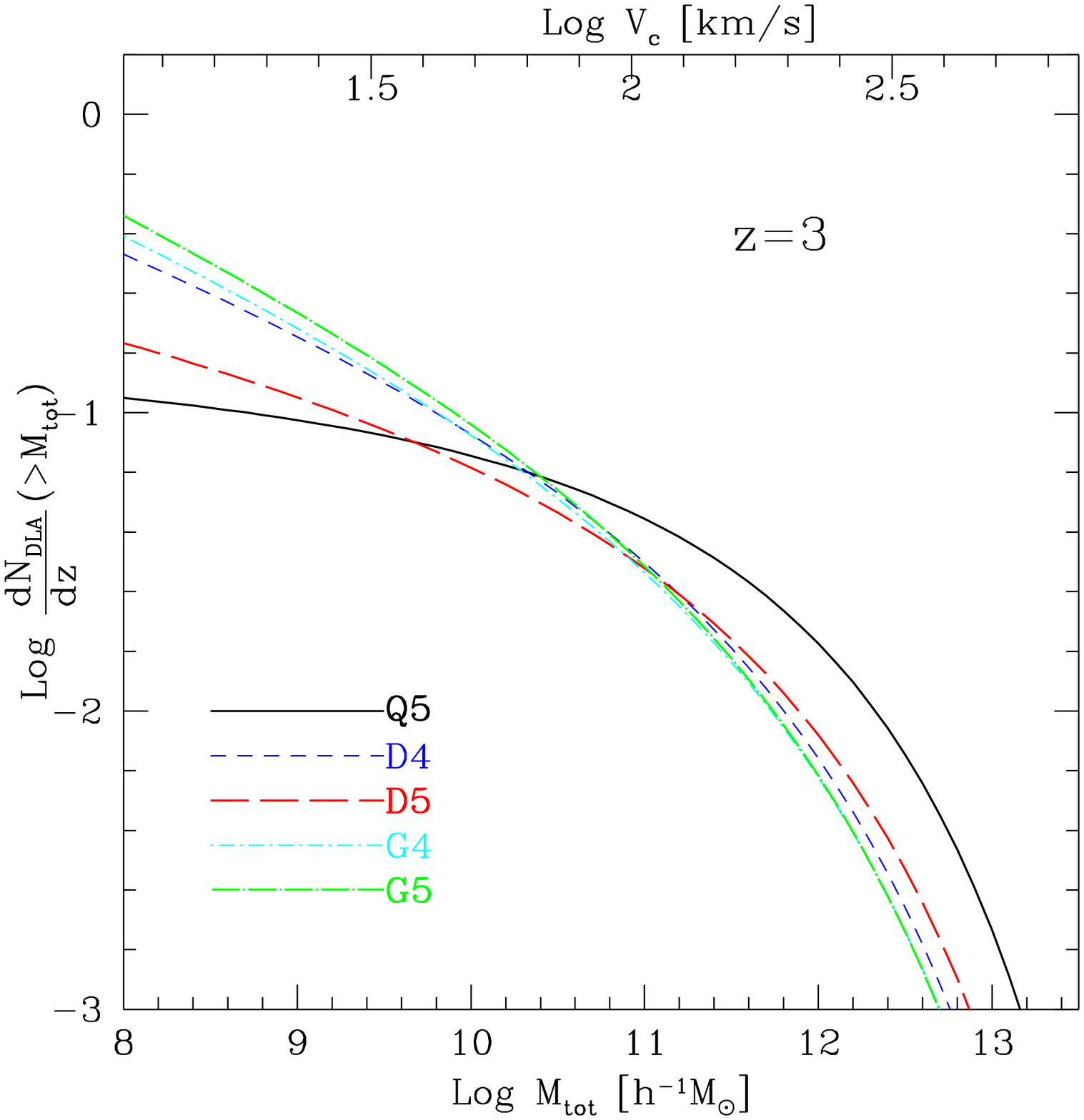}}\\%
\caption{Cumulative abundance (or equivalently rate of incidence) of DLAs 
per unit redshift as a function
of total halo mass. {\it Left panel:} Comparison of the results for
the Q-series. The runs with a strong wind (Q3, Q4, Q5) have fewer DLAs
than those with weak (P3) and no winds (O3).  The transition from Q3
to Q4, and then to Q5, shows that higher numerical resolution
tends to lead to fewer DLAs.  The shaded region indicates the observed
DLA abundance of \citet{Per01}.  The solid slanted line shows the
simulation result obtained by \citet{Gar01}.  {\it Right panel:} Same
as the left panel, but for the D- and the G-series, which all have
strong winds.  We also include the Q5 run as a thick solid line to ease
comparison with the left panel.
\label{cum_z3.eps}}
\end{center}
\end{figure*}

In Figure~\ref{cum_z3.eps}, we show the cumulative abundance of DLAs
per unit redshift at $z=3$ as a function of total halo mass. The
horizontal shaded region in the left panel indicates the observed DLA 
abundance of \citet{Per01}.  We note that the data-set analysed by 
\citet{Per01} includes that of \citet{Sto00}, and a similar value for
the 
DLA abundance was also reported by \citet{Sto00}.  It is encouraging 
that the DLA abundances found in our simulations agree well 
with the observed range.

As we discussed in Section~\ref{section:highz_area}, the DLA
cross-section is falling off rapidly at $\Mtot\sim 10^8\himsun$. 
Consequently,
the DLA abundance per unit redshift can be read off from the
cumulative abundance plot at $\Mtot=10^8\himsun$, provided that the
underlying simulation can resolve this mass-scale well.  For our
highest resolution run at $z=3$ (Q5), this is, in fact, the case. Here,
the cumulative abundance has already flattened out at
$\Mtot=10^8\himsun$, so that a correction with
equation~(\ref{eq:abundance}) for a missed contribution by haloes on
smaller mass-scales becomes unnecessary.

In the left panel of Figure~\ref{cum_z3.eps}, it is seen that the DLA
abundance decreases as the wind strength increases from O3 to P3, and
to Q3, and as the resolution of the simulation increases from Q3 to
Q4, and to Q5.  This is due to the increasing slope of the power-law
fits that we obtained in Section~\ref{section:cross}.  As the slope of
the power-law fit increases, the contribution from massive haloes
becomes larger, while that from low-mass haloes becomes smaller.  As a
result, the cumulative abundance at the high-mass end of
Figure~\ref{cum_z3.eps} is largest for Q5, but when summed over all
masses, Q5 exhibits the smallest total DLA abundance.

We also show the result from \citet{Gar01} as a solid slanted line in
the left panel of Figure~\ref{cum_z3.eps}.  They argued that their
result would be consistent with the observations if the observed DLAs
originate only from haloes with circular velocities larger than
$v_c\sim 60 \kms$ (which corresponds to $\Mtot\approx 2\times
10^{10}\himsun$).  However, the good agreement between our improved
simulations and the observational determinations suggests that
\citet{Gar01} probably overpredicted the DLA abundance due to the
shallower slope they estimated for the relation between the halo mass
and the DLA cross-section (see Section~\ref{section:z3_area}).

In the right panel of Figure~\ref{cum_z3.eps}, the results for the D-
and G-series are shown, with the values for Q5 included as a reference
to ease comparison with the left panel.  As the box-size increases
from the Q- to the D-series, and then to the G-series, the resolution
of the simulations severely degrades, causing an overprediction 
for the abundance, for the reasons discussed above.  Therefore, we 
believe that Q5 represents our most reliable estimate at $z=3$ (leaving 
aside the question whether the feedback strength of the galactic wind adopted 
for the Q-runs is appropriate or not).


\subsection{Redshift evolution of DLA abundance}
\label{section:abund_evolution}

\begin{figure*}
\begin{center}
\resizebox{8.2cm}{!}{\includegraphics{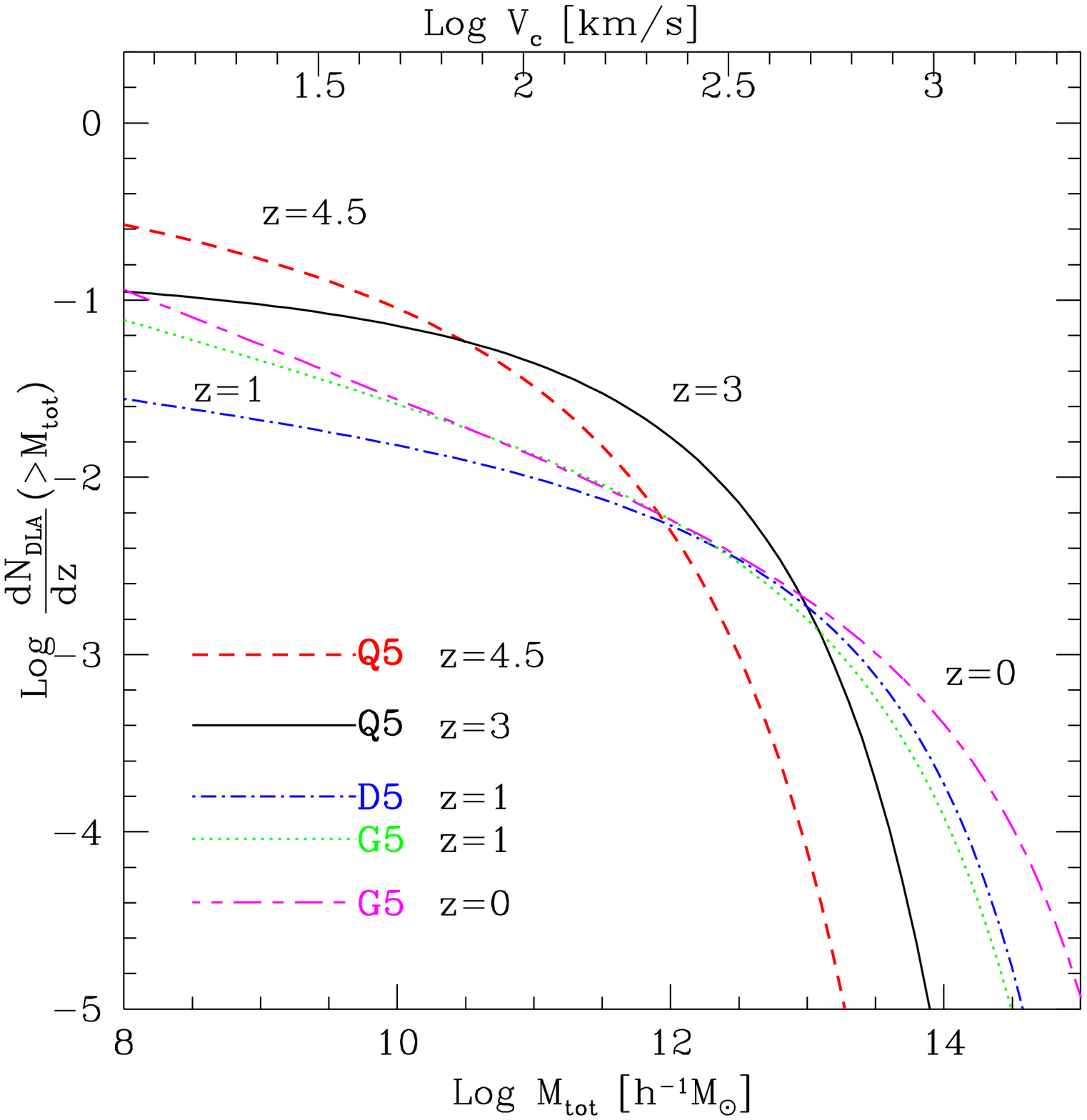}}%
\hspace{0.3cm}
\resizebox{8.2cm}{!}{\includegraphics{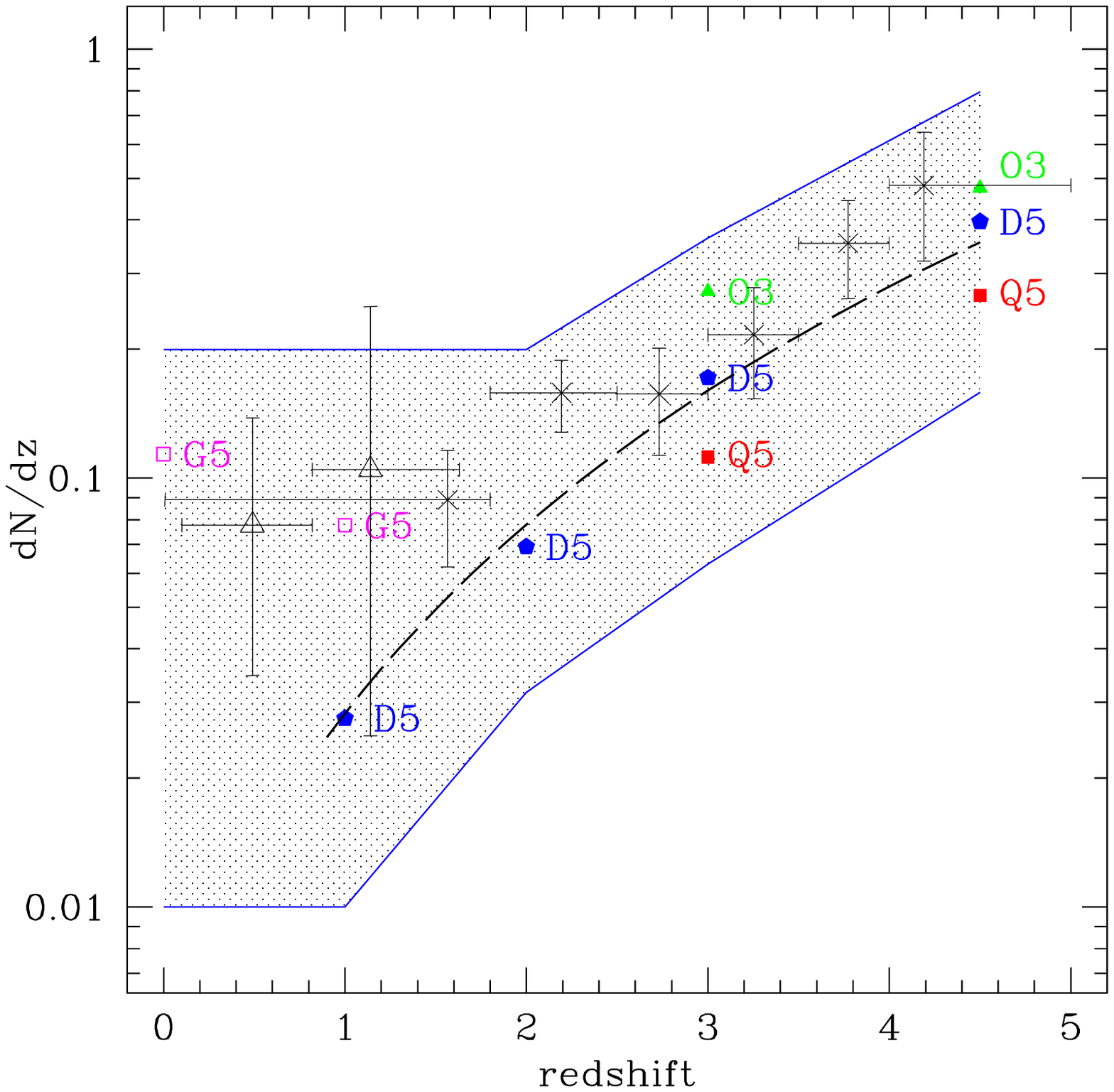}}\\%
\caption{Evolution of the DLA abundance from $z=4.5$ to $z=0$.  {\it
Left panel}: Cumulative DLA abundance as a function of total halo mass
at redshifts $z=4.5, 3, 1$ and 0.  {\it Right panel:} DLA abundance
per unit redshift as a function of redshift.  The data points with error bars
are the observational data from \citet{Per01} (crosses) and \citet{Rao00} 
(open triangles at $z<1.5$). The exact simulation results from some 
of the runs are indicated by the symbols with run-names.
The shaded region is our best-guess for a confidence region based on 
combining all of our simulations. For reference, we show the 
power-law of ${\rm d}N/{\rm d}z=N_0 (1+z)^\gamma$ with $N_0=0.005$ and $\gamma=2.5$
as a long-dashed line. 
\label{cum_evolve.eps}}
\end{center}
\end{figure*}

In Figure~\ref{cum_evolve.eps}, we show the evolution of DLA abundance
from $z=4.5$ to $z=0$.  In the left panel, the cumulative abundances
of DLAs are shown as a function of total halo mass for redshifts
$z=4.5, 3, 1$, and 0.  One can see that the contribution from massive
haloes to the DLA population progressively increases from high- to
low-redshift as a result of the merging of haloes.  In the right
panel, we give the DLA abundance per unit redshift as a function of
epoch, with values here read off at $\Mtot=10^8\himsun$ in the left
panel.  Observed data points from \citet{Per01} and \citet{Rao00} are
also shown with error bars. Some of the exact simulation results are
shown by the symbols, labelled with the names of the corresponding
runs.  The shaded region is our best-guess for a confidence region
based on combining all of our simulation results. For reference, we
show a power-law ${\rm d}N/{\rm d}z=N_0 (1+z)^\gamma$ with $N_0=0.005$ and
$\gamma=2.5$ as a long-dashed line, which describes the rate of
evolution seen in the simulations well.  It is encouraging that the
above value of $\gamma$ is in good agreement with the observed
evolution of Lyman-limit systems of \citet{Per01}, where they report
$\gamma=2.45^{+0.75}_{-0.04}$.

From $z=4.5$ to $z=3$, we see a decrease in the abundance by a factor
of about two in both simulations and the observations, and the agreement
between the two is very good, although the simulation points tend to
fall slightly below the observations.  From $z=3$ to $z=1$,
the simulation (D5) suggests a further rapid decrease in DLA abundance
by a factor of $\sim 6$, which is not seen at this level in the
existing observations.  But the observational data at
low redshift are still relatively uncertain, as indicated by the large
error bars.  The rapid decline is also reflected in the fact that
$\OHI$ decreases from 0.66 ($z=3$) to 0.14 ($z=1$) in D5 over this
redshift range, a reduction of nearly a factor of 5 (see
Figure~\ref{omega.eps}).  If this significant decrease in the number
of DLAs from $z=3$ to $z=1$ is real, it would partly explain why it is
so difficult to find DLAs at $z\leq 1$.

On the other hand, not much evolution is seen from $z=1$ to $z=0$ in
the G5 simulation.  This is related to the fact that $\OHI$ in G5 does
not decrease very much from $z=1$ ($\OHI=0.19$) to $z=0$
($\OHI=0.16$).  However, as discussed earlier, our power-law fits to
the $\sdla - {M_{\rm tot}}$ relation are not well constrained for
$z=0$ (and possibly for $z=1$ as well), so the results at $z\leq 2$
should be interpreted with caution.  At $z\geq 3$, we saw that lower
resolution runs tend to predict a larger abundance due to a shallower
slope in the relation between the DLA cross-section and the halo mass,
but it is not clear if other forms of systematic bias dominate at very
low redshift for simulations with poor resolution.  We will need yet
higher resolution simulations with large box-sizes to make a more
robust prediction of the DLA abundance at $z\leq 2$, and until then,
it is not clear whether the current results for DLA abundance at
$z\leq 2$, which tend to fall below the observational data, are
trustworthy. This is why we have widened the shaded confidence
region in Figure~\ref{cum_evolve.eps} significantly for $z\leq 2$.


\section{\HI Column Density Distribution Function}
\label{section:dist}

The column density distribution function $f(N,X(z))$ is defined such
that $f(N,X){\rm d}N{\rm d}X$ is the number of absorbers per sight
line with \HI column densities in the interval $[N,N+{\rm d}N]$, and
absorption distances in the interval $[X,X+{\rm d}X]$.  The absorption
distance $X(z)$ is given by \beq X(z) = \int_0^z (1+z')^2
\frac{H_0}{H(z')} {\rm d}z'.  \eeq This definition is based on an
argument by \citet{Bah69}, who pointed out that the probability of
absorption for a quasar sight-line in the redshift interval $[z,z+{\rm
d}z]$ is ${\rm d}P\propto (1+z)^2 {\rm d}r \propto (1+z)^2 [H_0 /H(z)]
{\rm d}z \equiv {\rm d}X$.  In practice, if the comoving box-size of
the simulation is $\Del L$, then the corresponding absorption distance
per sight-line is $\Del X = ({H_0}/{c})(1+z)^2 \Del L$.  For
example, for $\Del L=10\himpc$ and $z=3$, we have $\Del X= 0.0534$.

Assuming that DLAs do not overlap along a sight-line through the
simulation volume (which is a very good approximation given the small
size of the simulation box, where the expected number of DLAs per
sight-line at $z=3$ for a $10\himpc$ path is $\approx 10^{-3}$), we can
compute the $\NHI$ distribution function by counting the number of
grid-cells with column densities in the range $[N,N+{\rm d}N]$. In
doing so, we are treating each grid-cell element as one line-of-sight.


\subsection{\HI column density distribution at $z=3$}

\begin{figure*}
\epsfig{file=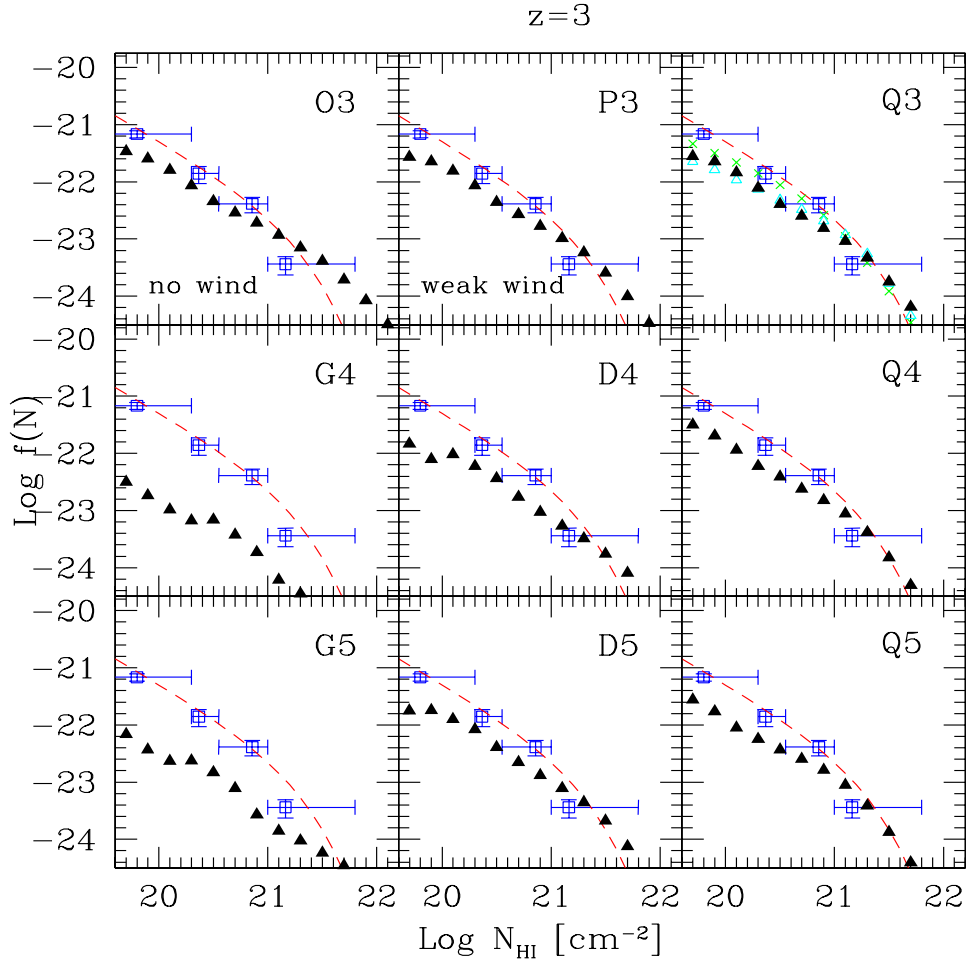,height=5in,width=5in, angle=0} 
\caption{ \HI column density distribution function at $z=3$.  The solid
triangles are the points measured directly from the simulations.
The open squares are the observational data of 
\citet[][for $2.7<z<3.5$ data]{Per01}, and the dashed line
is the fit to the same data based on a gamma-distribution.
In the panel for `Q3', the results with different smoothing methods are 
shown in crosses and open triangles. See text for details.
\label{dist_z3.eps}}
\end{figure*}

In Figure~\ref{dist_z3.eps}, we show the \HI column density distribution 
function at $z=3$.  The solid triangles are the points directly measured 
from the simulations. The open squares are the observational data of 
\citet[][for $2.7<z<3.5$ data]{Per01}, and the dashed curve is the fit 
to the same data based on a gamma-distribution:
\beq
\label{eq:gamma}
f(N)=\frac{f_\ast}{N_\ast}\left (\frac{N}{N_\ast}\right )^{-\beta}\exp\left(-\frac{N}{N_\ast}\right).
\eeq
The parameters of the fit are 
$(f_\ast, \log N_\ast, \beta)=(0.0406, 21.18, 1.10)$ 
\citet[][for $2.7<z<3.5$ data]{Per01}.
We note that all data by \citet{Sto00} are included in that of 
P\'eroux et al.'s.

In the panel for `Q3' (upper right corner), we also show the result of
different smoothing methods, using crosses (uniform cloud-in-cell
distribution with $\ell=[{4\pi}/{3}]^{1/3}s$) and open triangles
(uniform clouds-in-cell distribution with
$\ell=\frac{1}{2}[{4\pi}/{3}]^{1/3}s$). The former method (crosses)
results in higher values of $f(N)$ at lower column densities because
it smoothes the gas mass into broader regions. The SPH smoothing method
agrees with the latter calculation method (open triangles) better.

The agreement between the observations and the simulations Q3, Q4, Q5,
\& D5 at $\log\NHI>21$ is generally very good. Results from runs of
increasing resolution (Q3, Q4, and Q5) are consistent with each other
to a high degree.  The run with no wind (O3) somewhat overpredicts the
distribution function at large $\NHI$ values, but as the galactic wind
strength increases from O3 to P3, and then to Q3, the high column
density systems become less abundant and the agreement between the
simulation and observations improves.  At
intermediate column densities ($20<\log\NHI<21$), it seems that the 
simulated distribution function falls short of the observational estimate.  
Given the consistent behaviour in Q3, Q4, and Q5, our result
appears not to be affected by resolution, although this
cannot be completely excluded.  We will discuss this point further in
the next subsection, when we consider the data at $z=4.5$.  It is
clear however that G4 and G5 do not have sufficient resolution at
$z=3$ to resolve DLAs.


\subsection{\HI column density distribution at $z=4.5$}
\label{section:dist_z4.5}

\begin{figure*}
\epsfig{file=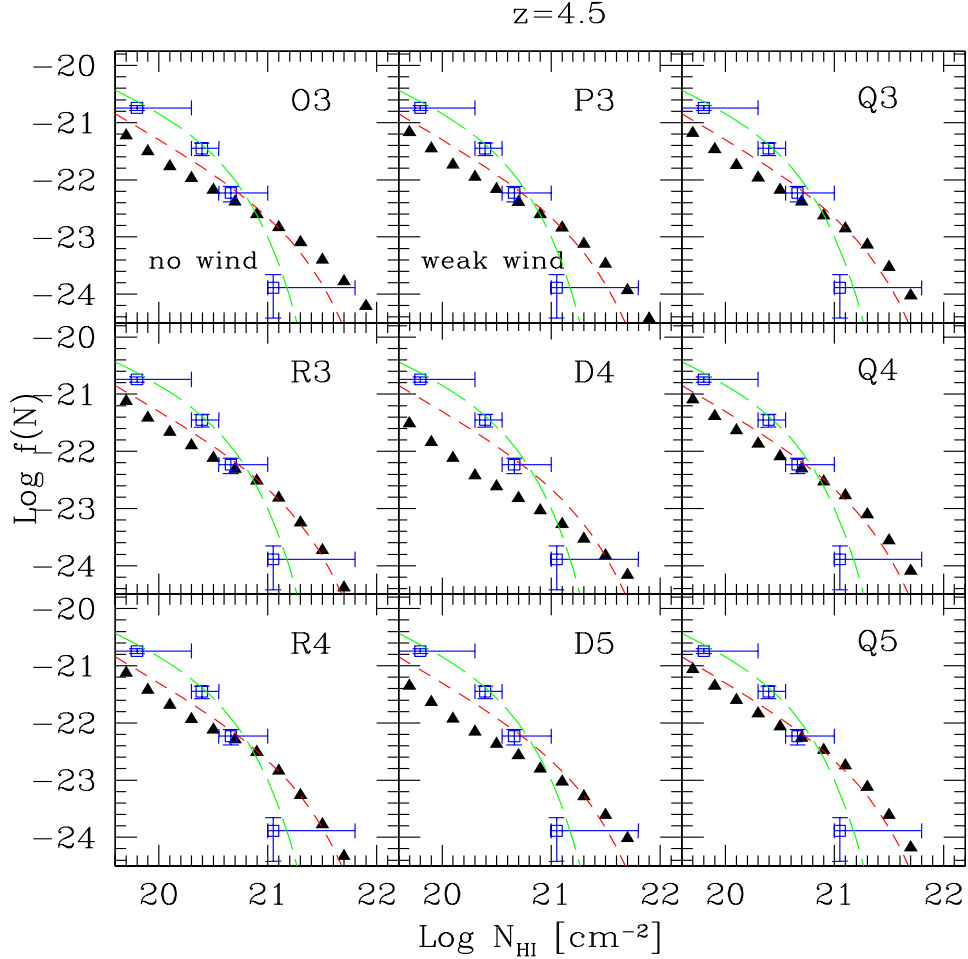,height=5in,width=5in, angle=0} 
\caption{\HI column density distribution function at $z=4.5$. 
The solid triangles are the points measured in the simulations, 
and the open squares are the observational data of 
\citet[][for $3.5<z<4.99$ data]{Per01}.
The long-dashed line is the gamma fit to the same observational 
data of $3.5<z<4.5$. For comparison, the short-dashed line is 
the fit to the $2.7<z<3.5$ data.
\label{dist_z4.5.eps}}
\end{figure*}

In Figure~\ref{dist_z4.5.eps}, we show the \HI column density 
distribution function at $z=4.5$. As before, the solid triangles 
are the points measured in the simulations, and the open squares 
are the observational data of \citet[][for $3.5<z<4.99$ data]{Per01}.
The long-dashed line is the gamma fit to the same observational 
data of $3.5<z<4.5$, and the short-dashed line is the fit to the 
data for $2.7<z<3.5$ for reference.
The values of the fit parameters for the $3.5<z<4.5$ data is 
$(f_\ast,\log N_\ast,\beta)=(0.2506, 20.46, 0.80)$.

Observational studies \citep{Sto00, Per01} indicate that there are
fewer high $\NHI$ systems ($\log\NHI>21$) at $z>3.5$ compared with
$2.7<z<3.5$, and that the distribution function becomes steeper at
$z>3.5$.  However, we do not see such a reduction of high $\NHI$
systems in our simulations from $z=3$ to $z=4$.  In fact, the highest
resolution simulation in our series (Q5) suggests that $f(N)$ is
slightly higher (but steeper at the same time) at $z=4.5$ compared 
to $z=3$.  Note that the agreement
between runs with different resolution (Q3, Q4, Q5, R3,
and R4) is impressive, showing that the results are well-converged.
The degree of the increase in $f(N)$ from $z=3$ to
$z=4.5$ is somewhat larger for the intermediate $\NHI$ systems
($20<\log\NHI<21$), leading to a slight steepening of the
overall $f(N)$.  One interpretation is that observational studies have
not yet discovered sufficient numbers of DLAs with very high column density
($\log\NHI>21$) to accurately estimate the evolution of $f(N)$ at
$z>3$, because such high $\NHI$ systems are intrinsically rare and
therefore difficult to find.

The evolution from $z=4.5$ to $z=3$
is presumably driven by the combined effect of the depletion of the
abundance of small haloes by merging, the accumulation of feedback by
galactic winds, and the reduction of cooling efficiency.  Small
haloes of comparatively low column densities ($\log\NHI<20.5$) merge
into larger systems from $z=4.5$ to $z=3$, forming more massive
systems with higher column densities, as the hierarchical structure
formation scenario suggests. However, strong feedback by galactic
winds ejects neutral gas from the star-forming regions, thereby acting
to reduce the column densities of all systems. In addition, the
efficiency of cooling rapidly decreases towards lower redshift,
reducing the rate at which gas can cool out of haloes and become
neutral.


\subsection{\HI column density distribution at lower redshift}
\label{section:dist_lowz}

\begin{figure}
\epsfig{file=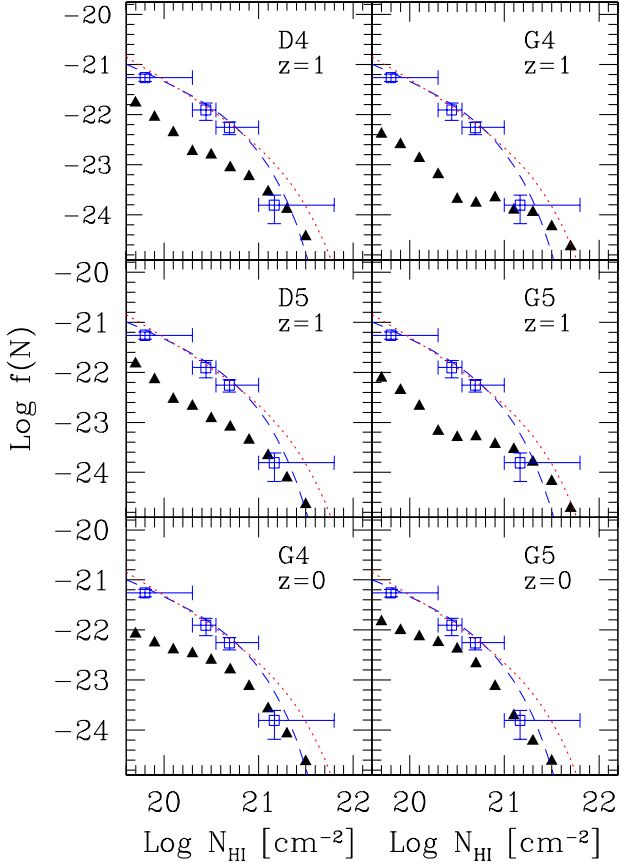,height=4.5in,width=3.2in, angle=0} 
\caption{\HI column density distribution function at $z=1$ and $z=0$.
The solid triangles are the points measured in the simulations,
and the open squares are the observational data of 
\citet[][for $0.008<z<2.0$ data]{Per01}.
The dashed curve is the gamma-distribution fit to the same observational 
data. For comparison, we also show the observational gamma fit for 
$2.7<z<3.5$ data as dotted lines.
\label{dist_lowz.eps}}
\end{figure}

In Figure~\ref{dist_lowz.eps}, we show the \HI column density distribution
function at $z=1$ and $z=0$.  Again, solid triangles give the points
measured in the simulations, and the open squares are the 
observational data of \citet[][for $0.008<z<2.0$ data]{Per01}. 
The dashed curve is the gamma-distribution fit to the same observational 
data with parameters $(f_\ast, \log N_\ast, \beta)=(0.0870, 20.76, 0.74)$.  
For comparison, we also show the observational gamma fit for 
$2.7<z<3.5$ data as dotted lines.

As we pointed out earlier, the results of the G-series are
substantially affected by resolution,
causing an underestimate of the DLA abundance if it is not corrected
using equation (\ref{eq:abundance}). (If the abundance is corrected
with an ill-determined relation between the DLA cross-section and the
halo mass, then low resolution can also lead to an overestimate of the
abundance, as we discussed in Section~\ref{section:abundance}.)  With
the mass resolution of the G-series, haloes with
$\Mtot<10^{10}\himsun$ are not resolved, and a significant fraction of
the DLA population is therefore missed.  Curiously, it is seen that
the column density distribution function in fact rises from $z=1$ to
$z=0$ in the G-series.  This is probably because haloes with masses
above the mass resolution limit are only beginning to form at these
low redshifts as the non-linear mass-scale increases with decreasing
redshift, and the simulation is finally ``catching up'' to reproduce
the neutral gas content in these haloes.

\section{Discussion}
\label{section:discussion}

We have used state-of-the-art hydrodynamic simulations of structure
formation to investigate the abundance of DLAs in a $\Lam$CDM
universe.  Our study represents a first attempt to apply a large
series of simulations to this problem, probing an unprecedented range
in both mass and spatial scales, enabling us to quantify systematic
effects due to numerical resolution.  Furthermore, we improved the
simulation methodology by adopting a novel formulation of SPH
\citep[see][]{SH02a} that minimises systematic inaccuracies in
simulations with cooling, and by using an improved model for the
treatment of the multiphase structure of the ISM in the context of
star formation and feedback \citep{SH02b}.

By comparing our results for DLA abundance in a series of runs as a
function of resolution and feedback strength, we were able to
demonstrate that insufficient resolution, or a lack of a proper
treatment of effective feedback processes, leads to an incorrect
estimate of the relation between the DLA cross-section and the halo
mass. This likely led to an overestimate of the DLA abundance in
earlier studies, for the reasons we discussed in detail in
Section~\ref{section:z3_area}.

\citet{Pro01} pointed out that the observed velocity width
distribution cannot be reproduced if the relation between the DLA
cross-section and the halo mass derived from these earlier SPH
simulations is used. They also suggested that one possibility to
remedy this inconsistency was to suppose that the relationship between
the DLA cross-section and the halo mass was incorrectly determined.
This is exactly what we find in our current study. If we use the new
relation found in our highest resolution simulation, we obtain a DLA
abundance that is consistent with observations, which is very
encouraging.  The slope of the relation that we infer from our 
simulations at $z=3$ is in the range of $0.7 - 1.0$, which coincides 
with the range that \citet{Hae00} derived by requiring their model
prediction of velocity width distribution to match the observed one.

However, while our simulations reproduce the DLA abundance at $z=3$ 
very well, our predictions at $z\leq 2$ are not equally reliable 
because they are based on simulations with larger box-sizes and lower 
resolution.  This is also evident from the poor agreement between the 
simulated and the observed \HI column density distribution function 
at low redshift.  To make the predictions at low-redshift more robust 
will require large box-size simulations with yet more particles.

The fact that we were able to reproduce the observed DLA abundance
very well at redshift $z=3$ has significant implications for the cold
dark matter model.  This result suggests that DLAs
arise naturally in a $\Lam$CDM universe from radiatively cooled gas in dark
matter haloes {\it with correct abundance}.  This is related to the
``substructure problem'' that is posed against the CDM model, based on
the notion that the number of satellite galaxies observed around the
Milky Way seems to be much smaller than, and hence in contradiction
to, the large number of dark matter substructures predicted by
high-resolution N-body simulations \citep{Moo99}. 

\citet{Stoehr02} have shown that the seriousness of this discrepancy
was overstated initially, but the high abundance of low mass
haloes and dark satellites in CDM models remains puzzling, given for
example the shallow faint-end slope of the galaxy luminosity function.
Therefore, many models have been invoked to suppress the formation of
dwarf satellite galaxies, including supernova feedback that ejects gas
\citep{Dek86}, reheating of the intergalactic medium which suppresses
gas infall \citep{Bul01}, and photoionisation of gas by UV background
radiation \citep{Tho96, Qui96}.  Full cosmological hydro-simulations
have also been used to argue that feedback effects by UV background
radiation and supernovae may account for the suppression of the
formation of low-mass galaxies relative to the steeply rising dark
matter halo mass function \citep{Chi01, Nag01b}.  While we cannot make
a strong statement on this problem at low-redshift, our results
suggest that the CDM model does not have difficulty at $z=3$ with
respect to the number of neutral gas clumps from which stars
form. Considering that there is in general good agreement between
observations and theoretical studies of Lyman-break galaxies at $z=3$
\citep[e.g.][]{Mo96, Bau98, Jin98, KHW99, Kau99, Mo99, Nag02, Wei02}, 
the $\Lam$CDM model seems to be in a very good shape at $z=3$.

Another interesting result of our study is that the break in the
relation between DLA cross-section and halo mass seems to occur at a
mass-scale of $\Mtot\approx 10^8 - 10^{8.5}\himsun$.  DLAs do not
exist in haloes below this mass-scale at $z=3$ and $z=4.5$. Because we
measured this effect consistently in both the `Q5' and `R4'-runs,
which have sufficient resolution to describe haloes with
$\Mtot=10^8\himsun$ well, it is clear that this is not caused by a
resolution effect.  Note that this mass-scale, which corresponds to
circular velocities of $10-15\kms$ and virial temperatures of about
$10^4\,{\rm K}$ or even slightly lower, is smaller than suggested by
the earlier studies of \citet{Qui96} and \citet{Tho96} (See also 
\citet{Wei97}).  The physical origin of the break could lie in both 
the radiative processes of cooling and UV-heating, and in supernovae 
feedback, but we expect that the former effects dominate.  This is 
because feedback will proceed only if at least some neutral gas is 
built up, so that star formation can occur at some (low) level.  
Only {\it after} that can feedback quench the further accumulation 
of neutral gas, and possibly also affect neighbouring haloes. 
Therefore, while one can easily understand why supernovae feedback 
reduces the amount of neutral gas in small halos, it is not clear 
how it could lead to the complete absence of neutral gas below a 
certain mass-scale. On the other hand, a relatively sharp break can 
be expected from the properties of the cooling function, and the 
heating of the gas due to the UV background. Both processes depend 
very sensitively on the virial temperature of halos in the range 
$10^4 - 10^5$\,{\rm K}.  Further studies will be needed to 
disentangle the relative importance of the various physical effects 
in this regime more precisely.

Assuming that feedback effects due to star formation nevertheless
influence the mass-scale of the break, we might have underestimated it
if the feedback we employed in our simulations was not strong
enough. However, we saw in our results of Figure~\ref{omega.eps} that
the current simulations with the `strong wind' model already seem to
underpredict $\OHI$ slightly, therefore even stronger feedback
would make the neutral gas fraction in the Q-series even smaller and
exacerbate the discrepancy with the observational estimate of $\OHI$
at $z=3$.  Also, there is only very limited room to increase the
strength of the UV background flux, because for our chosen
normalisation the mean opacity of the Lyman-$\alpha$ forest is in
good agreement with observations.  Another way of constraining the
feedback strength is to compare the simulated galaxy luminosity
function with the observed one, to see if the flatness of the
faint-end of the luminosity function can be reproduced.  A preliminary
analysis suggests that our present simulations do
not yet yield a luminosity function with a faint-end as flat as
observed at low-redshift; therefore the feedback model in our current
simulation might not be adequate to correctly account for the galaxy
population at $z=0$.  The results of this analysis will be
presented elsewhere. A solution to this problem will help to further
constrain the physical processes that suppress the DLA cross-section
in low-mass haloes.

In the present paper, we focused on the abundance of DLAs, without
carrying out a detailed study of their internal structure.
Therefore, we are not able to completely rule out the possibility that
the DLAs we find in our simulations are in fact originating from very
small disks at the center of dark matter haloes.
Our present work is complementary to that of Haehnelt et al. (1998),
in the sense that we are able to sample a large statistical ensemble 
of absorbers in comoving volume of $\geq (10\himpc)^3$, while they 
studied a smaller number of objects in greater detail.
Nevertheless, our highest resolution run (Q5) with $10\himpc$ box 
has a spatial resolution of $1.2\hikpc$ which is comparable to that of 
Haehnelt et al. (1998), and the results from varying resolution runs 
(Q3, Q4, Q5) agree with each other well.  Therefore, given that our 
simulations are based on the CDM model, our results support the idea 
that DLAs arise from radiatively cooled protogalactic gas clumps 
embedded in dark matter haloes.
While our spatial resolution is approaching comoving sub-kpc scales 
at $z=3$, it is probably not yet sufficient to accurately resolve 
the disk structures embedded in the central few kpc regions of dark 
matter haloes.  Kinematical studies of the internal structure of 
DLAs will require yet higher resolution simulations than those used 
in the present paper, and possibly a more sophisticated physical model 
for the ISM and star formation.


\section*{Acknowledgements}

We thank Celine P{\'e}roux for providing us with the data points in
Figures~\ref{omega.eps}, \ref{cum_evolve.eps}, \ref{dist_z3.eps}, 
\ref{dist_z4.5.eps}, \& \ref{dist_lowz.eps}. We are also grateful to 
Art Wolfe, Jason Prochaska, and Eric Gawiser for useful discussions.  
This work was supported in part by NSF grants ACI 96-19019, AST 98-02568, 
AST 99-00877, and AST 00-71019.  The simulations were performed at the
Center for Parallel Astrophysical Computing at the Harvard-Smithsonian
Center for Astrophysics.


\bsp

\label{lastpage}

\end{document}